\documentclass[%
reprint,
amsmath,amssymb,
aps,
prc,
floatfix,
twocolumn,
twoside,
showkeys,
showpacs
]{revtex4-2}

\usepackage[T1]{fontenc}
\usepackage{graphicx}
\usepackage{siunitx}
\usepackage{lmodern}
\usepackage{newtxtext,newtxmath}
\usepackage{hyperref}
\hypersetup{
    colorlinks=true,
    linkcolor=blue,
    citecolor=blue,
    urlcolor=blue,
    filecolor=blue
}

\usepackage{fancyhdr}
\fancypagestyle{plain}{%
  \fancyhf{}
  \fancyfoot[C]{\thepage}
  
}

\usepackage{microtype}
\setcitestyle{numbers,square,comma,sort&compress}

\newcommand{\asympt}[2]{\;\raisebox{-2mm}{$\widetilde{\scriptstyle{#1\to #2}}$}\;}

\begin{document}

\title{Numerical study of the two-boson bound-state problem with and without
partial-wave decomposition}
\author{Wolfgang Schadow}
\email{\!\!wolfgang.schadow@caribou3d.com}
\affiliation{
  Caribou3D Research \& Development \\
53424 Remagen, Kirchplatz 1, Germany
}

\date{April 26, 2026}

\begin{abstract}
The validation of numerical methods is a prerequisite for reliable few-body
calculations, particularly when moving beyond standard partial-wave
decompositions. In this work, we present a precision benchmark for the two-boson
bound-state problem, solving it using two complementary formulations:
the standard one-dimensional partial-wave Lippmann--Schwinger equation and a
two-dimensional formulation based directly on vector variables. While the
partial-wave approach is computationally efficient for low-energy bound
states, the vector-variable formulation becomes essential for scattering applications
at higher energies where the partial-wave expansion converges slowly. We
demonstrate the high-precision numerical equivalence of both methods using
rank-one separable Yamaguchi potentials and non-separable Malfliet--Tjon
interactions. Furthermore, for the Yamaguchi potential, we derive exact
analytical expressions quantifying the systematic errors introduced by finite
momentum- and coordinate-space cut-offs. These analytical bounds provide a
rigorous tool for disentangling discretization errors from truncation effects
in few-body codes. The results establish a highly controlled methodological
benchmark that provides a detailed baseline for vector-variable algorithms
intended for more complex three- and four-body calculations.
\end{abstract}


\maketitle
\thispagestyle{plain}

\renewcommand{\arraystretch}{1.02}

\section{Introduction}

The description of systems of three or more interacting particles, as
encountered in atomic and nuclear physics, requires the solution of the Faddeev
or Faddeev--Yakubovsky equations \cite{Elster98a,Elster98b,Schadow99d,
Polasek2000,Shertzer2001,Caia2004,Kessler2004,Kadyrov2005,Liu2005a,Ramalho2006,
Liu2007a,Hadizadeh2007a,RodriguezGallardo2008,Hadizadeh2008a,Bayegan2008b,
Harzchi2010a,Gloeckle2010a,Golak2010a,Golak2012a,Shalchi2012a,Veerasamy2013,
Harzchi2014a,Kuruoglu2016}. Their numerical
implementation is a computationally demanding task, for which reliable
verification and precise error control are essential. A well-established
strategy for validating such calculations is their application to simpler
subsystems for which high-precision or analytical results are available.

The two-boson bound-state problem serves as a useful benchmark in this
context. Although it represents the simplest bound-state system, it already
contains key features of the interaction and the underlying quantum mechanics.
The binding energy and wave function can therefore be used to test numerical
accuracy, quadrature schemes, and the treatment of singular behavior in a
controlled setting.

While the partial-wave decomposition (PWD) is the standard method for solving
the Schrödinger equation at low energies, it becomes computationally burdensome
in scattering calculations at intermediate and high energies due to the slow
convergence of the partial-wave series. In such regimes, formulations based
directly on vector variables, which avoid the angular momentum expansion, are
advantageous \cite{Elster98a,Schadow99d,Liu2005a,Golak2010a}. The 3D non-PWD
numerical solution of the two-body Lippmann--Schwinger equation for the off-shell
$T$-matrix has been extensively studied, particularly with regard to multivariable
quadrature and convergence issues\cite{Elster98b, Polasek2000, Shertzer2001, Caia2004,
Kessler2004, Kadyrov2005, Ramalho2006, RodriguezGallardo2008, Veerasamy2013,
Kuruoglu2016}. The present study does not constitute a validation of non-PWD
methods for the off-shell two-body $T$-matrix; rather, it provides a
controlled preliminary benchmark for the underlying bound-state quadrature,
interpolation, and eigenvalue machinery. In contrast to the earlier two-body
non-PWD scattering studies, the present work focuses on constructing a
precision benchmark for bound-state implementations, including exact
analytical cutoff-error formulas for the Yamaguchi model and a detailed
comparison of 1D and 2D numerical realizations at the $10^{-10}$~MeV level.

In this work, we construct such a benchmark by solving the Lippmann--Schwinger
equation for the two-boson ground state in two complementary ways. The first is
the conventional approach based on a partial-wave decomposition, which leads to
a set of uncoupled one-dimensional integral equations. The second avoids a
partial-wave expansion and treats the equation directly as a two-dimensional
integral equation in momentum magnitude and angle between the momenta. This
formulation is closer to the structure encountered in three-body calculations
and allows for a direct comparison of numerical performance.

To demonstrate the robustness of the methods, we consider two classes of
interactions. The rank-one separable Yamaguchi potential
\cite{Yamaguchi54a,Yamaguchi54b} allows analytical solutions and enables a
direct comparison with exact results. In addition, it permits the derivation of
closed expressions for the systematic errors introduced by momentum- and
coordinate-space cut-offs. As a more realistic example, we study local
Malfliet--Tjon potentials \cite{Malfliet69}, which are of Yukawa type
\cite{Yukawa35a} and pose different numerical challenges due to their
short-range structure. The two-body equations cannot be solved in simple closed
form; only approximate methods \cite{Sabet2021,Hamzavi2012} or analytical
expansions based on supersymmetry \cite{Napsuciale2021} are available.

This paper is organized as follows. In Sec.~\ref{sec:partial}, we derive the
partial-wave-projected Lippmann--Schwinger equation for the two-boson bound
state. Section~\ref{sec:twobwithoutpartial} presents the corresponding
two-dimensional formulation. The applications to the Yamaguchi and
Malfliet--Tjon potentials are discussed in Secs.~\ref{sec:boundyamaguchi}
and~\ref{sec:boundmt}. The numerical methods are described in
Sec.~\ref{sec:numerical_method}. Results and convergence studies are presented
in Sec.~\ref{sec:results}. Section~\ref{sec:summary} summarizes the work. The
appendices collect analytical expressions for expectation values of the
Yamaguchi potential in momentum and coordinate space.

The specific contributions of this work beyond previous studies include
the derivation of exact analytical expressions for the systematic errors
induced by finite momentum- and coordinate-space cut-offs for the Yamaguchi
potential, and a precision benchmark of the vector-variable formulation
achieving $10^{-10}$~MeV consistency with partial-wave results, thereby
providing a controlled benchmark for future high-precision applications.

\section{Partial-wave-projected Lippmann--Schwinger equation}
\label{sec:partial}

We consider the two-boson bound-state problem in momentum space and set
$\hbar = c = 1$. The mass of each boson is chosen as $m_N = 1$, which implies
a reduced mass $\mu = 1/2$. The binding energy $E$ and the bound-state wave
function $|\psi\rangle$ are determined from the Schrödinger equation
\begin{equation}
H \, |\psi\rangle = (H_0 + V) \, |\psi\rangle = E \, |\psi\rangle \,,
\label{eqschroedinger}
\end{equation}
where we work in the two-body center-of-mass frame. The free Hamiltonian is
given by $H_0 = p^2$, and $V$ denotes the two-body interaction. Note that with
these definitions, the energy $E$ has dimensions of inverse length squared
(fm$^{-2}$). Physical energies in MeV are obtained by multiplication with the
conversion factor $(\hbar c)^2/m_N$.

Equation~(\ref{eqschroedinger}) can be written in Lippmann--Schwinger form as
\begin{equation}
|\psi\rangle = \frac{1}{E - H_0} \, V \, |\psi\rangle \,.
\label{eq:lippmann}
\end{equation}
For bound states, the energy is negative. Throughout this work, we denote the
two-body binding energy as $E_2 < 0$. For the Yamaguchi potential, we specifically
write $E_2 = -\alpha^2$. Since $E_2 < 0$, the denominator $E_2 - p^2$ in the
integral equation does not vanish for any real momentum $p \ge 0$.

To solve this equation, we project onto momentum states and introduce
partial-wave states $|p\,l\,m\rangle$. These states satisfy the completeness
relation
\begin{equation}
1 = \sum_{l m} \int\limits_0^\infty \!dp\, p^2\, |p\,l\,m\rangle \langle
  p\,l\,m| \,,
\label{eq:complete1d}
\end{equation}
and are normalized according to
\begin{equation}
\langle p\,l\,m | p'\,l'\,m' \rangle =
\delta_{l l'}\,\delta_{m m'}\,\frac{\delta(p-p')}{p^2} \,.
\end{equation}
Here, $p$ denotes the relative momentum, $l$ the orbital angular momentum, and
$m$ the corresponding magnetic quantum number.

After projection with $\langle p\,l\,m|$ and insertion of the completeness
relation~(\ref{eq:complete1d}), Eq.~(\ref{eq:lippmann}) becomes
\begin{align}
\psi_{lm}(p) \equiv \langle p\,l\,m | \psi \rangle = &
\sum_{l' m'} \int\limits_0^\infty \!dp'\, p'^2\,
\frac{1}{E - p^2} \nonumber \\
& \times \langle p\,l\,m | V | p'\,l'\,m' \rangle\,
\psi_{l'm'}(p') \,.
\label{eqpsi}
\end{align}

We restrict our study to spherically symmetric interactions without tensor
forces. In this case, the matrix element of the potential is diagonal in the
angular momentum quantum numbers,
\begin{equation}
\langle p\,l\,m | V | p'\,l'\,m' \rangle =
\delta_{l l'}\,\delta_{m m'}\, V_l(p,p') \,,
\end{equation}
with the partial-wave-decomposed potential given by
\begin{equation}
V_l(p,p') =
\int\limits_{-1}^{1} \!d(\cos\theta)\, P_l(\cos\theta)\, V({\mathbf p},{\mathbf p}') \,.
\end{equation}
Here, $\theta$ denotes the angle between the relative momentum vectors ${\mathbf p}$
and ${\mathbf p}'$. As a result, Eq.~(\ref{eqpsi}) reduces to
\begin{equation}
\psi_l(p) =
\int\limits_0^\infty \!dp'\, p'^2\,
\frac{1}{E - p^2}\,
V_l(p,p')\,\psi_l(p') \,,
\label{eq:lspsinotensor}
\end{equation}
which is the one-dimensional Lippmann--Schwinger equation. Since the equation is
independent of $m$, the magnetic quantum number has been omitted.

For the interactions considered here, only the $l=0$ partial wave supports a
bound state. We therefore restrict the discussion to this channel and drop the
angular momentum indices for the wave function. The wave function is normalized
according to
\begin{equation}
1 = \langle \psi | \psi \rangle =
\int\limits_0^\infty \! dp\, p^2\, |\psi(p)|^2 \,.
\label{eq:normpsi1d}
\end{equation}
The expectation values of the kinetic and potential energy operators are given
by
\begin{equation}
\langle \psi | H_0 | \psi \rangle =
\int\limits_0^\infty \!dp\, p^2\, \psi^*(p)\, p^2\, \psi(p) \,,
\end{equation}
and
\begin{equation}
\langle \psi | V | \psi \rangle =
\int\limits_0^\infty \!dp\, p^2 \int\limits_0^\infty \! dp'\, p'^2\,
\psi^*(p)\, V_0(p,p')\, \psi(p') \,.
\end{equation}

\section{Two-body bound-state equation without partial-wave decomposition}
\label{sec:twobwithoutpartial}

In this approach, the Lippmann--Schwinger equation~(\ref{eq:lippmann}) is
projected onto vector momentum states $|{\mathbf p}\rangle$. These states satisfy
the completeness relation
\begin{equation}
1 = \int \! d^3p \, |{\mathbf p}\rangle \langle {\mathbf p}| \,,
\label{eq:complete2d}
\end{equation}
and are normalized according to
\begin{equation}
\langle {\mathbf p} | {\mathbf p}' \rangle
= \delta({\mathbf p}-{\mathbf p}')
= \frac{\delta(p-p')}{p^2}\, \delta(\hat{\mathbf{p}}-\hat{\mathbf{p}}') \,.
\end{equation}
Here $\hat{\mathbf{p}}$ denotes the unit vector specified by the angular variables
$\Omega_p=(\theta,\varphi)$. The Dirac delta distribution on the unit sphere is
defined by its action
\begin{equation}
\int \! d\Omega_{\hat p'} \,
\delta(\hat{\mathbf{p}}-\hat{\mathbf{p}}')\, f(\hat{\mathbf{p}}') = f(\hat{\mathbf{p}}),
\end{equation}
for any test function $f(\hat{\mathbf{p}})$, where
$d\Omega_{\hat p}=d(\cos\theta)\,d\varphi$. In spherical coordinates $(\theta,\varphi)$,
this distribution can be written explicitly as
\begin{equation}
\delta(\hat{\mathbf{p}}-\hat{\mathbf{p}}')
\equiv \delta(\Omega_p-\Omega_{p'})
= \frac{\delta(\theta-\theta')}{\sin\theta}\,\delta(\varphi-\varphi').
\end{equation}

After projection with $\langle {\mathbf p} |$ and insertion of the completeness
relation~(\ref{eq:complete2d}), Eq.~(\ref{eq:lippmann}) takes the form
\begin{equation}
\langle {\mathbf p} | \psi \rangle
= \int \! d^3p' \, \frac{1}{E-p^2}\,
\langle {\mathbf p} | V | {\mathbf p}' \rangle
\langle {\mathbf p}' | \psi \rangle \,.
\label{eq:ls3d}
\end{equation}

We restrict ourselves to interactions for which the matrix element $\langle {\bf
p} | V | {\mathbf p}' \rangle$ is a rotationally invariant scalar function,
\begin{equation}
\langle {\mathbf p} | V | {\mathbf p}' \rangle
= V({\mathbf p},{\mathbf p}')
= V(p,p',\hat{\mathbf{p}}\cdot\hat{\mathbf{p}}') \,,
\end{equation}
where $\hat{\mathbf{p}}\cdot\hat{\mathbf{p}}'$ is the cosine of the angle between ${\bf
p}$ and ${\mathbf p}'$.

Introducing the variables $x = \hat{\mathbf{z}}\cdot\hat{\mathbf{p}}$ and
$x' = \hat{\mathbf{z}}\cdot\hat{\mathbf{p}}'$, and denoting the azimuthal
angles of $\hat{\mathbf p}$ and $\hat{\mathbf p}'$ by $\varphi$ and $\varphi'$, respectively,
Eq.~(\ref{eq:ls3d}) can be written as
\begin{align}
\psi(p,x)
& = \int\limits_0^\infty \!dp'\, p'^2 \int\limits_{-1}^{1} \! dx'
\int\limits_0^{2\pi} \!d\varphi \, \frac{1}{E-p^2}\nonumber \\
& \qquad \qquad \times  V(p,p',y)\, \psi(p',x') \,,
\label{eq:ls2dfullv}
\end{align}
where
\begin{equation}
y = \hat{\mathbf{p}}\cdot\hat{\mathbf{p}}'
= x \, x' + \sqrt{1-x^2}\, \sqrt{1-x'^2}\, \cos(\varphi-\varphi') \,.
\end{equation}
Without loss of generality, the azimuthal angle of $\hat{\mathbf{p}}$ is set to
zero.

Defining
\begin{align}
  \label{eq:vppxx}
 &v(p,p',x,x')  \\
&\qquad= \int\limits_0^{2\pi} \!d\varphi\,
V\!\left(p,p',x \, x' + \sqrt{1-x^2} \, \sqrt{1-x'^2}\cos\varphi \right), \nonumber
\end{align}
the integral equation (\ref{eq:ls2dfullv}) reduces to
\begin{align}
\psi(p,x)
&= \int\limits_0^\infty \!dp'\, p'^2 \int\limits_{-1}^{1} \!dx'\,
\frac{1}{E-p^2}\, \nonumber \\
& \qquad \qquad \times v(p,p',x,x')\, \psi(p',x') \,.
\label{eq:ls2d}
\end{align}
This is a two-dimensional integral equation in the magnitude $p$ and the angular
variable $x$ (hereafter referred to as the 2D formulation).

Although Eq.~(\ref{eq:vppxx}) allows the use of the full interaction directly,
it is instructive to expand the potential in Legendre polynomials
\begin{equation}
V(p,p',\hat{\mathbf{p}}\cdot\hat{\mathbf{p}}')
= \sum_l \frac{2l+1}{4\pi}\,
P_l(\hat{\mathbf{p}}\cdot\hat{\mathbf{p}}')\, V_l(p,p') \,.
\end{equation}
Inserting this expansion into Eq.~(\ref{eq:vppxx}) yields
\begin{align}
v(p,p',x,x')
&= \sum_l \frac{2l+1}{4\pi}\, V_l(p,p')
\int\limits_0^{2\pi} \!d\varphi\, P_l(y) \nonumber \\
&= \sum_l \frac{2l+1}{2}\, V_l(p,p')\, P_l(x)\, P_l(x') \,,
\label{eq:vppxxpart}
\end{align}
which allows the use of partial-wave-decomposed interactions within the
two-dimensional formulation.

To demonstrate the equivalence of the one- and two-dimensional approaches, the
wave function is expanded as
\begin{equation}
\psi(p,x)
= \sum_l \sqrt{\frac{2l+1}{4\pi}}\, \psi_l(p)\, P_l(x) \,.
\label{eq:psi2dpartial}
\end{equation}
Substituting Eqs.~(\ref{eq:psi2dpartial}) and~(\ref{eq:vppxxpart}) into
Eq.~(\ref{eq:ls2d}), using the orthogonality of the Legendre polynomials, and
integrating over $x$, one obtains
\begin{equation}
\psi_l(p)
= \frac{1}{E-p^2} \int\limits_0^\infty \!dp'\, p'^2\,
V_l(p,p')\, \psi_l(p') \,.
\label{eq:uncoupled_system}
\end{equation}
The two-dimensional equation (\ref{eq:ls2d}) is therefore equivalent to an infinite set of
uncoupled one-dimensional equations, one for each partial wave.

For the interactions considered here, only the $l=0$ channel is sufficiently
attractive to support a bound state. All higher partial waves are suppressed by
the centrifugal barrier and do not support negative-energy solutions. As a
result, the bound-state solution of the full two-dimensional equation must be a
pure $s$-wave, and its binding energy must coincide with that obtained from the
one-dimensional $l=0$ equation. Verifying this equivalence numerically provides
a central validation of the two-dimensional formulation.

The normalization condition reads
\begin{equation}
1 = \langle \psi | \psi \rangle
= 2\pi \int\limits_0^\infty \!dp\, p^2 \int\limits_{-1}^{1} \!dx\, |\psi(p,x)|^2 \,,
\end{equation}
where the factor $2\pi$ arises from the analytical integration over the azimuthal
angle $\varphi$. The expectation values of the kinetic and potential energy operators
are given by
\begin{equation}
\langle \psi | H_0 | \psi \rangle
= 2\pi \int\limits_0^\infty \!dp\, p^2 \int\limits_{-1}^{1} \!dx\,
\psi^*(p,x)\, p^2\, \psi(p,x) \,,
\end{equation}
and
\begin{align}
\langle \psi | V | \psi \rangle
=& 2\pi \int\limits_0^\infty \!dp\, p^2  \int\limits_{-1}^{1} \!dx
\int\limits_0^\infty \!dp'\, p'^2 \int\limits_{-1}^{1} \!dx'\,\nonumber \\
&\qquad
\times \psi^*(p,x)\, v(p,p',x,x')\, \psi(p',x') \,,
\end{align}
where $v(p,p',x,x')$ denotes the potential after the integration over the
azimuthal angle $\varphi$ has been carried out, see~Eq.~(\ref{eq:vppxx}).

\begin{table*}[htbp]
\caption{
\linespread{1.3}\selectfont
Parameters and conversion factors for the Yamaguchi potentials. The resulting
two-body binding energies $E_2$ are calculated using Eq.~(\ref{eq:energyyamaguchi}).
Listed values are the exact ones used in calculations. Parameter sets YAMA-0 and
YAMA-23, and YAMA-96 are defined in this work.}
\label{tab:yamaguchiparams}
\begin{ruledtabular}
\setlength{\tabcolsep}{0pt}
\begin{tabular*}{\textwidth}{
    @{\extracolsep{\fill}}
    l
    S[table-format=1.7, group-digits=false]
    S[table-format=1.7, group-digits=false]
    S[table-format=2.8, group-digits=false]
    S[table-format=3.7, group-digits=false]
    S[table-format=-1.12, group-digits=false]
    c
}
Potential &
\multicolumn{1}{c}{$\beta$} &
\multicolumn{1}{c}{$\lambda$} &
\multicolumn{1}{c}{${(\hbar c)^2}/{m_N}$} &
\multicolumn{1}{c}{$\hbar c$} &
\multicolumn{1}{c}{$E_2$} &
Ref. \\
 & {[fm$^{-1}$]} & {[fm$^{-3}$]} & {[MeV fm$^2$]} & {[MeV fm]} & {[MeV]} & \\
\hline\\[-8pt]
YAMA-0   & 1.14525   & 0.22      & 41.47106096 & 197.327053  & -2.225936306578 & \\
YAMA-23  & 1.3905818 & 0.3707654 & 41.47103997 & 197.3269804 & -2.224567197074 & \\
YAMA-96  & 1.15      & 0.184     & 41.47106096 & 197.327053  & -0.471586629956 & \\
YAMA-IV  & 1.15      & 0.179     & 41.47       & 197.3286    & -0.331781328055 & \cite{Hadizadeh2007a}
\end{tabular*}
\end{ruledtabular}
\\[-2pt]
\end{table*}

\section{Two-body bound-state equation for the Yamaguchi potential}
\label{sec:boundyamaguchi}

\subsection{Wave function in momentum space}

A rank-one separable potential has the general form
\begin{equation}
V = - |g\rangle \, 4\pi\lambda \, \langle g| \,,
\label{eq:yamaguchipot}
\end{equation}
where $|g\rangle$ is the form factor and $\lambda$ is the strength parameter.
Inserting Eq.~(\ref{eq:yamaguchipot}) into the Lippmann--Schwinger
equation~(\ref{eq:lippmann}) yields, in momentum space,
\begin{equation}
|\psi\rangle
= -\frac{1}{E-p^2}\, |g\rangle \, 4\pi\lambda \, \langle g|\psi\rangle
= N \, \frac{1}{\alpha^2+p^2}\, |g\rangle \,,
\label{eq:psisepform}
\end{equation}
with the scalar quantity
\begin{equation}
N = 4\pi\lambda \, \langle g|\psi\rangle \,,
\quad\text{and}\quad
E = -\alpha^2 \,.
\label{eq:norm}
\end{equation}
Substituting Eq.~(\ref{eq:psisepform}) into Eq.~(\ref{eq:norm}) and assuming
$N\neq 0$ yields an implicit relation between $\lambda$ and the binding energy,
\begin{equation}
\lambda^{-1}
= 4\pi \, \langle g | \frac{1}{\alpha^2+p^2} | g \rangle \,.
\label{eq:lambda}
\end{equation}

Depending on the explicit form of $|g\rangle$, Eq.~(\ref{eq:lambda}) can be
evaluated analytically. We employ generalized form factors defined for all
partial waves,
\begin{equation}
\langle p\, l | g\rangle \equiv g_l(p)
= \frac{p^l}{(p^2+\beta^2)^{l+1}} \,.
\label{eq:yamaguchiform}
\end{equation}
For $l=0$ this reduces to the original Yamaguchi form factor. A bound state
exists only in the $l=0$ channel for the parameter sets considered here.
Inserting $g_0(p)$ into Eq.~(\ref{eq:lambda}) gives
\begin{equation}
\alpha = \sqrt{\frac{\pi^2 \, \lambda}{\beta}} - \beta \,,
\quad\text{provided that} \quad
\lambda > \frac{\beta^3}{\pi^2} \,,
\label{eq:energyyamaguchi}
\end{equation}
and the normalized two-body wave function follows from
Eqs.~(\ref{eq:normpsi1d}) and~(\ref{eq:psisepform}),
\begin{equation}
\psi(p)
= 2 \, \sqrt{\frac{\alpha \, \beta \,(\alpha+\beta)^3}{\pi}}\,
\frac{1}{\alpha^2+p^2}\, \frac{1}{p^2+\beta^2} \,.
\label{eq:psiyamaguchipspace}
\end{equation}

Using Eq.~(\ref{eq:psiyamaguchipspace}), the expectation value of the
Hamiltonian can be expressed analytically in terms of $\alpha$, $\beta$, and
$\lambda$,
\begin{align}
 \langle \psi | H| \psi \rangle
 & =  \langle \psi | H_0 | \psi \rangle  +  \langle \psi | V| \psi \rangle \\
&=  \bigg( \alpha \,\beta \bigg) + \bigg(  - \frac{\pi^2 \, \alpha \, \lambda}
{\beta \,(\alpha + \beta)} \bigg)   = -\alpha^2 \,.
\end{align}

For the error analysis, it is useful to express the exact expectation values as
limits of finite momentum cut-offs. We define
\begin{align}
\langle H_0 \rangle_{\rm cut}
&\equiv \int\limits_{0}^{p_{\rm cut}} \! dp \, p^2 \, \psi^*(p)\, p^2\, \psi(p)\,,  \\
\langle \psi | H_0 | \psi \rangle
&= \lim_{p_{\rm cut}\to\infty} \langle H_0 \rangle_{\rm cut}\,,
\intertext{and}
\langle V \rangle_{\rm cut}
&\equiv -4\pi\lambda
\left(\int\limits_{0}^{p_{\rm cut}} \!dp \, p^2 \, \psi(p)\,
\frac{1}{p^2+\beta^2}\right)^{\!\!2},  \\
\langle \psi | V | \psi \rangle
&= \lim_{p_{\rm cut}\to\infty} \langle V \rangle_{\rm cut}\,.
\end{align}
For the Yamaguchi potential with the form factor of
Eq.~(\ref{eq:yamaguchiform}), these expressions can be evaluated analytically.
The resulting closed forms are given in Appendix~\ref{sec:limityamaguchi}. They
provide quantitative upper bounds on the accuracy achievable at finite $p_{\rm
cut}$ and guide the choice of mesh resolution.

In the literature, several parameter sets are used for the Yamaguchi potential,
often together with different conversion factors. Table~\ref{tab:yamaguchiparams}
lists the parameter sets considered here, the conversion factors employed, and
the exact two-body binding energy $E_2$ obtained from
Eq.~(\ref{eq:energyyamaguchi}).

\subsection{Wave function in coordinate space}

The coordinate-space wave function is obtained by Fourier transformation of the
momen\-tum-space wave function,
\begin{equation}
\psi(r)
= \sqrt{\frac{2}{\pi}} \int\limits_{0}^{\infty} \!dp \, p^2 \, \psi(p)\, j_0(p\,r)\,,
\label{eq:psifourier}
\end{equation}
where $j_0(p \,r)=\sin(p \, r)/(p \,r)$ is the spherical Bessel function. For the Yamaguchi
form, the integral can be evaluated analytically. Using
Eq.~(\ref{eq:psiyamaguchipspace}), one finds
\begin{equation}
\psi(r)
= \sqrt{\frac{2\,\alpha \, \beta \,(\alpha+\beta)}{(\alpha-\beta)^2}}\,
\frac{e^{-\alpha \, r}-e^{-\beta \, r}}{r}\,.
\label{eq:psirspace}
\end{equation}
This is the normalized $l=0$ solution of the Schrödinger equation for a
separable, non-local interaction in coordinate space,
\begin{align}
-\frac{1}{r^2}\frac{\partial}{\partial r}
\left(r^2\frac{\partial}{\partial r}\psi(r)\right) &  -4\pi\lambda\,
g(r)\int\limits_{0}^{\infty} \!dr' \, r'^2 \, g(r')\, \psi(r')
\nonumber \\
&= E\, \psi(r)\,,
\label{eq:schroedingerrspace}
\end{align}
where the Fourier-transformed interaction reads
\begin{equation}
V(r,r') = - g(r)\, 4\pi\lambda\, g(r')\,,
\end{equation}
with
\begin{equation}
g(r) = \sqrt{\frac{\pi}{2}} \, \frac{e^{-\beta \, r}}{r}\,.
\end{equation}
It is convenient to write $\psi(r)=u(r)/r$. Then
Eq.~(\ref{eq:schroedingerrspace}) becomes
\begin{equation}
-\frac{\partial^2}{\partial r^2}u(r)
-4\pi\lambda\, g(r)\int\limits_{0}^{\infty} \! dr' \, r' \, g(r')\, u(r')
= E\, u(r)\,,
\end{equation}
and the normalization condition for $\psi(r)$ reduces to
\begin{equation}
1 = \langle \psi | \psi \rangle = \int\limits_{0}^{\infty}\! dr \, |u(r)|^2\,.
\end{equation}
For any interaction that falls off faster than $1/r^2$ as $r\to\infty$, the
bound-state wave function behaves as
\begin{equation}
\psi(r) \asympt{r}{\infty} A_s \, \frac{e^{-\alpha  \, r}}{r}\,,
\end{equation}
where $A_s$ is the asymptotic normalization constant. As $\beta>\alpha$ here,
the asymptotic form of Eq.~(\ref{eq:psirspace}) is
\begin{equation}
\psi(r) \asympt{r}{\infty}
\sqrt{\frac{2\,\alpha \, \beta \,(\alpha+\beta)}{(\alpha-\beta)^2}}\,
\frac{e^{-\alpha \, r}}{r}\,.
\end{equation}
In this case, $A_s$ coincides with the normalization constant in
Eq.~(\ref{eq:psirspace}). With the explicit form of Eq.~(\ref{eq:psirspace}),
the expectation value of the radius can be evaluated analytically as
\begin{equation}
\langle \psi | \! \left( \frac{1}{2} \,r \right) \!| \psi \rangle
= \frac{1}{4}
\left(\frac{1}{\alpha}+\frac{1}{\beta}+\frac{2}{\alpha+\beta}\right)\!,
\end{equation}
and the rms radius is
\begin{align}
&\sqrt{\langle \psi | \! \left(\frac{1}{2} \,r\right)^{\!2} \! | \psi
\rangle}
  \nonumber \\
& \qquad = \sqrt{\frac{1}{8}  \left(\frac{1}{\alpha^2}+\frac{1}{\beta^2}
+\frac{3}{\alpha \, \beta}+\frac{4}{(\alpha+\beta)^2}\right)}\;.
\end{align}
These exact results can again be written as limits of finite coordinate
cut-offs $r_{\rm cut}$ via the expressions in
Eqs.~(\ref{eq:limitradius}) and~(\ref{eq:limitrmsradius}). They provide upper
bounds on the accuracy achievable at finite $r_{\rm cut}$ and guide the choice
of mesh resolution.

\section{Two-body bound-state equation for the Malfliet--Tjon potential}
\label{sec:boundmt}

As a second interaction, we consider local Malfliet--Tjon potentials of Yukawa
type \cite{Yukawa35a,Malfliet69}. While these potentials do not describe
realistic nuclear forces with tensor components and spin-dependence, they
possess a strong short-range repulsion (hard core) characteristic of realistic
interactions. This feature makes them an ideal testing ground for numerical
algorithms, as the strong repulsion generates significant high-momentum
components in the wave function, posing a stringent test for the stability of
the integral equation solvers and the cut-off convergence. In coordinate space,
they are given by
\begin{equation}
V(r) = V_a \, \frac{e^{-\mu_a \, r}}{r} + V_r \, \frac{e^{-\mu_r \, r}}{r}\,,
\label{eqmalfliet-tjon}
\end{equation}
where the long-range attractive part is characterized by the strength $V_a$ and
the range parameter $\mu_a$. The short-range repulsive part is described by the
strength $V_r$ and the range parameter $\mu_r$. For a local potential of this
form, the partial-wave matrix elements in momentum space can be obtained from
the Fourier transformation
\begin{equation}
V_l(p,p')
= \frac{2}{\pi}\int\limits_{0}^{\infty} \! dr \, r^2 \, j_l(p \,r)\, V(r)\, j_l(p'\,r)\,.
\label{vlpprimep}
\end{equation}

\begin{table*}[htbp]
\caption{
\linespread{1.3}\selectfont
Parameters for the Malfliet--Tjon potentials and conversion factors from
the cited references. Listed values are the exact ones used in calculations.}
\label{tab:mtparams}
\begin{ruledtabular}
\setlength{\tabcolsep}{0pt}
\sisetup{exponent-product = \cdot}
\begin{tabular*}{\textwidth}{
    @{\extracolsep{\fill}}
    l
    S[table-format=-3.4, group-digits=false]
    S[table-format=1.3, group-digits=false]
    S[table-format=4.4, group-digits=false]
    S[table-format=1.2, group-digits=false]
    S[table-format=2.2, group-digits=false]
    S[table-format=3.4, group-digits=false]
    c
}
Potential &
\multicolumn{1}{c}{$V_{a}$} &
\multicolumn{1}{c}{$\mu_{a}$} &
\multicolumn{1}{c}{$V_{r}$} &
\multicolumn{1}{c}{$\mu_{r}$} &
\multicolumn{1}{c}{${(\hbar c)^2}/{m_N}$} &
\multicolumn{1}{c}{$\hbar c$} &
Ref. \\
 & {[MeV fm]} & {[fm$^{-1}$]} & {[MeV fm]} & {[fm$^{-1}$]} & {[MeV fm$^2$]} & {[MeV fm]} & \\
\hline\\[-8pt]
MT-III  & -635.306  & 1.55  & 1458.047  & 3.11 & 41.47 & 197.3    & \cite{Malfliet69} \\
MT-IV   & -65.109   & 0.633 & {--}      & {--} & 41.47 & 197.3    & \cite{Elster98b}\\
MT-V    & -570.3316 & 1.55  & 1438.4812 & 3.11 & 41.47 & 197.3    & \cite{Elster98b}\\
MT-VI   & -58.7954  & 0.723 & {--}      & {--} & 41.47 & 197.3    & \cite{Malfliet69} \\
\end{tabular*}
\end{ruledtabular}
\\[-2pt]
\end{table*}

Inserting Eq.~(\ref{eqmalfliet-tjon}) into Eq.~(\ref{vlpprimep}) yields the
closed form
\begin{equation}
V_l(p,p') = \frac{1}{\pi}\left( \frac{V_a}{p\,p'}\, Q_l(z_a)
+ \frac{V_r}{p\,p'}\, Q_l(z_r)
\right),
\end{equation}
with
\begin{equation}
z_{a/r} = \frac{p^2 + p'^2 + \mu_{a/r}^2}{2 \,p \,p'} \,,
\end{equation}
where $Q_l(z)$ denotes the Legendre functions of the second kind for $|z|>1$. The
first few functions are
\begin{alignat}{2}
Q_0(z) & = \frac{1}{2}\, \ln\!\left(\frac{z+1}{z-1}\right), \\
Q_1(z) & = \frac{z}{2}\, \ln\!\left(\frac{z+1}{z-1}\right) - 1, \\
Q_2(z) & = \frac{1}{4}\, (3z^2-1) \, \ln\!\left(\frac{z+1}{z-1}\right) - \frac{3}{2} \, z \,.
\end{alignat}
For the numerical treatment, the potential matrix elements are required at the
momentum boundaries where $p$ or $p'$ vanishes. There is only a contribution
for $l = 0$ in this limit
\begin{align}
&V_0(p,p') \bigg|_{p \cdot p' = 0} \nonumber \\
& \qquad \qquad= \frac{2}{\pi} \left( \frac{V_a}{p^2 + p'^2 + \mu_a^2}
 + \frac{V_r}{p^2 + p'^2 + \mu_r^2} \right) .
\end{align}

For the formulation without partial-wave decomposition, we require the
momentum-space matrix element $\langle{\mathbf p}|V|{\mathbf p}'\rangle$. For a local
interaction $\langle{\mathbf r}|V|{\mathbf r}'\rangle = V(r)\,\delta^{(3)}({\mathbf r}-
{\mathbf r}')$, one has
\begin{equation}
\langle{\mathbf p}|V|{\mathbf p}'\rangle
= \frac{1}{(2\pi)^3}\int \! d^3r \,
e^{-i\,({\mathbf p}-{\mathbf p}')\cdot{\mathbf r}}\, V(r)\,.
\end{equation}
Evaluating this integral for Eq.~(\ref{eqmalfliet-tjon}) gives
\begin{align}
& V({\mathbf p},{\mathbf p}') \nonumber \\
 &\qquad = \frac{1}{2\pi^2}\left(
  \frac{V_a}{({\mathbf p}-{\mathbf p}')^2+\mu_a^2} + \frac{V_r}{({\mathbf p}-
  {\mathbf p}')^2+\mu_r^2} \right).
\end{align}
With this explicit representation, the azimuthal integral in Eq.~(\ref{eq:vppxx})
can also be evaluated analytically. One finds
\begin{widetext}
\begin{align}
v(p,p',x,x') &= \frac{1}{\pi}\left[
\frac{V_a}{\sqrt{
\left(p^2+p'^2-2 \,p \, p' \, x \, x' + \mu_a^2\right)^2
-4 \, p^2 \,p'^2(1-x^2)(1-x'^2)}}
\right. \nonumber \\
&\qquad\qquad\qquad\qquad \left.
+ \frac{V_r}{\sqrt{
\left(p^2+p'^2-2 \, p \, p' \, x \, x' + \mu_r^2\right)^2
-4 \, p^2 \,p'^2(1-x^2)(1-x'^2)}}
\right].
\end{align}
\end{widetext}
In the literature, several parameter sets for the Malfliet--Tjon potentials are
used, often together with different conversion constants. Table~\ref{tab:mtparams}
collects the parameter sets adopted here and the corresponding conversion
factors.

\section{Numerical Methods}
\label{sec:numerical_method}

The determination of the two-body bound-state energy and wave function in
momentum space requires the solution of the homogeneous Lippmann--Schwinger
equation (\ref{eq:lspsinotensor}) or (\ref{eq:ls2d}) for negative energies.
Depending on the interaction, this can be achieved analytically or by
transforming the integral equations into matrix eigenvalue problems.

\subsection{Special Case: The Yamaguchi Potential}

The separable nature of the Yamaguchi potential allows for more direct methods
to find the binding energy than the general matrix eigenvalue approach. First,
it can be calculated exactly using the analytical formula given in
Eq.~(\ref{eq:energyyamaguchi}). With $E = -\alpha^2$, this equation provides the
exact binding energy directly from the potential parameters $\lambda$ and
$\beta$.

Alternatively, the binding energy can be found by numerically solving the
implicit relation in Eq.~(\ref{eq:lambda})
\begin{equation}
\lambda^{-1} =  4\pi \,\langle g | \frac{1}{E - p^2} | g \rangle =
 4\pi  \int\limits_0^\infty \! dp \, p^2 \, \frac{g^2(p) }{E - p^2} \,.
\label{eq:lambda_integral}
\end{equation}
To evaluate the integral over the semi-infinite domain, we map the standard
Gauss--Legendre interval $x \in (-1, 1)$ to the momentum range $p \in (0,
\infty)$ using the transformations
\begin{equation}
p(x) = s \, \frac{1+x}{1-x} \qquad {\text{and}} \qquad dp = \frac{2
\,s}{(1-x)^2} \, dx \,,
\end{equation}
where $s$ is a scaling factor to optimize the point distribution. For the
parameter sets considered here, the integrand in Eq.~(\ref{eq:lambda_integral})
peaks in the momentum range $0.25$--$0.5$~fm$^{-1}$. Choosing $s = 1$~fm$^{-1}$
maps the interval $x \in (-1, 0)$ to $p \in (0, 1)$, ensuring that the peak
region is resolved with high density, while the interval $x \in (0, 1)$ covers
the high-momentum tail. The binding energy $E_2$ is then obtained by solving
this equation numerically with a standard root-finding procedure.

\subsection{Discretization of the Integral Equations}

For a general potential, the 1D partial-wave-projected Lippmann--Schwinger
equation (\ref{eq:lspsinotensor}) is a homogeneous integral equation. To solve
this numerically, we discretize the integral over the continuous variable $p'$
using Gauss--Legendre grid points. For the numerical treatment, the
semi-infinite integration domain is truncated at a cut-off momentum $p_{\rm
cut}$. To ensure a sufficient density of quadrature points in the physically
important low-momentum region, the integration interval $(0, p_{\rm cut})$ is
subdivided into two parts, $(0, p_0) \cup (p_0, p_{\rm cut})$, with $N_{P1}$ and
$N_{P2}$ points respectively. For the precise distribution of the quadrature
points within each sub-interval, we employ the mapping detailed in
Ref.~\cite{Gloeckle91}. To allow for the evaluation of the wave function at the
origin, the point $p=0$ with weight zero is explicitly added to the mesh. Since
both the Green's function $(E-p^2)^{-1}$ (for $E<0$) and the potential matrix
elements considered here remain finite at $p=0$, this inclusion does not
introduce numerical instabilities. This results in a final momentum mesh of
$N_P = N_{P1} + N_{P2} + 1$ points.

For the 1D partial-wave equation (\ref{eq:lspsinotensor}), this discretization
leads to a system of linear equations:
\begin{equation}
\psi(p_i) = \frac{1}{E - p_i^2} \, \sum_{j=1}^{N_{P}} \left[ wp_j \, p_j^2 \,
V_l(p_i, p_j) \right] \psi(p_j) \,.
\end{equation}
This represents a matrix eigenvalue problem $\vec{\psi} = K(E) \, \vec{\psi}$,
where the kernel matrix $K(E)$ has dimension $N_{P} \times {N}_{P}$ with
elements
\begin{equation}
K_{ij}(E) = \frac{1}{E - p_i^2} \, wp_j \, p_j^2 \, V_l(p_i, p_j) \,,
\end{equation}
where $wp_j$ are the Gauss--Legendre weights. The method used to determine the
solution is described in Sec.~\ref{sec:eigenvalue}.

To solve the 2D Lippmann--Schwinger equation (\ref{eq:ls2d}), we must discretize
two continuous variables. We again use Gauss--Legendre quadrature for both
integrations. For the angular variable $x$, the standard Gauss--Legendre points
lie within the open interval $(-1, 1)$. To avoid numerical extrapolation in
subsequent applications, we explicitly include the boundary points $x = \pm 1$
in the grid. These auxiliary points are assigned zero weights to
preserve the exactness of the quadrature rule. Consequently, the full 2D grid
for $\psi(p_i, x_k)$ includes the boundaries $p=0$ and $x=\pm 1$, ensuring the
wave function is defined over the entire closed domain. Applying these
quadrature rules to Eq.~(\ref{eq:ls2d}), we obtain
\begin{align}
\psi(p_i, x_k) = & \frac{1}{E - p_i^2} \, \sum_{j=1}^{{N}_{P}} \sum_{l=1}^{N_{X}}
\left[ wp_j \, p_j^2 \, wx_l \right .\nonumber \\
& \qquad \qquad\times \left. v(p_i, p_j, x_k, x_l) \right] \psi(p_j, x_l) \,.
\label{eq:discretized_2d}
\end{align}
To obtain a standard matrix eigenvalue problem, the two-dimensional wave
function $\psi(p_i,x_k)$ is rewritten as a vector of dimension ${N} = {N}_{P}
\times {N}_{X}$. A composite index $I = (i,k)$ maps each grid point to a single
vector index, such that $\psi_I = \psi(p_i, x_k)$. The discretized equation
again takes the compact form $\vec{\psi} = K(E) \, \vec{\psi}$, where the kernel
matrix $K(E)$ is of size $N \times N$, and its elements are given by
\begin{align}
K_{IJ}(E) = \frac{1}{E - p_i^2} \, \left[ wp_j \, p_j^2 \, wx_l \,
v(p_i, p_j, x_k, x_l) \right] ,
\label{eq:kernel_matrix_2d}
\end{align}
with $I=(i,k)$ and $J=(j,l)$.

\subsection{Solution of the Eigenvalue Problem}
\label{sec:eigenvalue}

Both the 1D and 2D approaches yield a homogeneous equation of the form
$K(E) \, \vec{\psi} = \lambda(E) \, \vec{\psi}$. The physical binding energy
$E_2$ corresponds to the condition $\lambda(E_2) = 1$. It should be noted that
for an isolated two-body problem with an energy-independent potential, it is
computationally more direct and efficient to cast the Schrödinger equation as
a symmetric matrix eigenvalue problem in momentum space \cite{Karr2010}.
However, the present approach of searching for the unit eigenvalue of the
non-symmetric Lippmann--Schwinger kernel $K(E)$ is chosen deliberately. In
the context of three- and four-body Faddeev or Faddeev--Yakubovsky equations,
the effective interaction kernels are inherently energy-dependent and
non-symmetric. By solving the two-body problem via this unit-eigenvalue
search, we rigorously test the exact iterative eigenvalue routines and
non-symmetric matrix handling required for these higher-dimensional applications.

The kernel matrix $K(E)$ is generally non-symmetric and dense. Furthermore,
the presence of strong short-range repulsion in potentials like the
Malfliet--Tjon interaction leads to large negative eigenvalues in the
spectrum of $K$. Since standard power iteration converges to the eigenvalue
with the largest magnitude (often a mode driven by the repulsive core with a
large negative value), it cannot be used directly to find the physical bound
state near $\lambda=1$.

Traditionally, inverse iteration with a shift was employed to isolate the
physical state~\cite{Wilkinson65, Gloeckle82b}. However, this requires an
$O(N^3)$ matrix factorization (such as LU decomposition) at every energy step,
which becomes prohibitively expensive for the large matrices encountered in
the 2D vector-variable formulation.

In this work, we employ the Restarted Arnoldi Method (RAM)~\cite{Arnoldi51},
a Krylov subspace technique. Stadler \textit{et al.}~\cite{Stadler91a}
pioneered the application of such iterative methods to few-body integral
equations to avoid the storage and inversion of large kernel matrices. The
algorithm proceeds by constructing an orthogonal basis $V_m$ for the Krylov
subspace $\mathcal{K}_m(K, \vec{v}) = \text{span}\{\vec{v}, K\vec{v}, \dots,
K^{m-1}\vec{v}\}$ of dimension $m \ll N$. The projection of the large matrix
$K$ onto this subspace yields a small $m \times m$ upper Hessenberg matrix
$H_m = V_m^T K V_m$. The eigenvalues of $H_m$, usually obtained via the QR
algorithm, provide excellent approximations to the extreme eigenvalues of $K$.

To ensure convergence to the physical state and minimize memory usage, we use
an iterative restarting strategy. If the approximate eigenvector obtained from
the subspace of size $m$ (typically $m \approx 40$) does not satisfy the
convergence criterion, the method is restarted using this vector as the new
initial guess. This process effectively filters out the unphysical components
of the spectrum. Since the physical bound state corresponds to the
algebraically largest positive eigenvalue of $K$ (distinct from the large
negative eigenvalues arising from the repulsive core), the method converges
rapidly to the desired solution.

The dominant computational cost in this approach is the matrix-vector
multiplication, which scales as $O(N^2)$, in contrast to the $O(N^3)$ scaling
of direct inversion methods. This efficiency gain is crucial for the 2D
formulation, where $N$ can reach $\mathcal{O}(10^4)$ for dense momentum grids.

\section{Results and Discussion}
\label{sec:results}

\subsection{The Yamaguchi Potential: Benchmarking and Validation}

The primary advantage of the Yamaguchi potential is its analytical solvability,
which provides an exact benchmark for all numerical methods. In
Table~\ref{tab:yamaguchiparams}, we list the exact binding energies $E_2$ for
various parameter sets, calculated directly from the analytical expression in
Eq.~(\ref{eq:energyyamaguchi}). These values serve as the benchmark against
which the numerical results are measured.

\begin{table}[htbp]
\caption{
\linespread{1.3}\selectfont
Convergence of the two-body binding energy $E_2$ for YAMA-23 and YAMA-IV,
obtained by numerically solving the implicit Eq.~(\ref{eq:lambda_integral}).
$N_P$ denotes the number of Gauss--Legendre integration points.}
\label{tab:implicit_convergence}
\begin{ruledtabular}
\setlength{\tabcolsep}{0pt}
\begin{tabular*}{\textwidth}{
    @{\extracolsep{\fill}}
    S[table-format=2.0]
    S[table-format=-1.12, group-digits=false]
    S[table-format=-1.12, group-digits=false]
}
\multicolumn{1}{c}{$N_P$} &
\multicolumn{1}{c}{YAMA-23} &
\multicolumn{1}{c}{YAMA-IV} \\
 &
\multicolumn{1}{c}{$E_2$ [MeV]} &
\multicolumn{1}{c}{$E_2$ [MeV]} \\
\hline\\[-8pt]
 8 & -2.221497086914 & -0.333671149537 \\
16 & -2.224567143808 & -0.331789172180 \\
24 & -2.224567197077 & -0.331781342438 \\
32 & -2.224567197074 & -0.331781328071 \\
40 & -2.224567197074 & -0.331781328055 \\
48 & -2.224567197074 & -0.331781328055 \\
\end{tabular*}
\end{ruledtabular}
\\[-2pt]
\end{table}

A first numerical test is to solve the implicit
equation~(\ref{eq:lambda_integral}) for the binding energy. The results for two
representative parameter sets, YAMA-23 and YAMA-IV, are shown in
Table~\ref{tab:implicit_convergence}. These examples demonstrate that as the
number of quadrature points $N$ increases, the calculated binding energy rapidly
converges to the exact analytical value. This behavior is typical for all
parameter sets investigated. Typically, 32 quadrature points are sufficient to
match the analytical binding energy to 12 significant digits. All potentials
reach this level of accuracy by $N=40$, confirming the high precision of the
mapping and quadrature schemes.

\begin{table*}[!htbp]
\caption{
\linespread{1.3}\selectfont
Momentum-space convergence for the YAMA-23 potential using the analytical
wave function given in Eq.~(\ref{eq:psiyamaguchipspace}). Deviations
$\Delta X \equiv X_{\text{exact}} - X_{\text{cut}}$ and
$\Delta E \equiv E_2 - (\langle H_0\rangle_{\rm cut}+\langle V\rangle_{\rm cut})$
are relative to the exact limits $\langle H_0\rangle=13.35647233$~MeV,
$\langle V\rangle=-15.58103952$~MeV, and $E_2=-2.22456720$~MeV. $N_P$ is the
number of Gauss--Legendre points.}
\label{tab:momentum_convergence_YAMA23}
\begin{ruledtabular}
\setlength{\tabcolsep}{0pt}
\sisetup{exponent-product = \cdot}
\begin{tabular*}{\textwidth}{
    @{\extracolsep{\fill}}
    S[table-format=4.0, group-digits=false]
    S[table-format=2.8, group-digits=false]
    S[table-format=1.2e-2, group-digits=false]
    S[table-format=-2.8, group-digits=false]
    S[table-format=-1.2e-1, group-digits=false]
    S[table-format=-1.8, group-digits=false]
    S[table-format=-1.2e-2, group-digits=false]
    S[table-format=3.0, group-digits=false]
}
\multicolumn{1}{c}{$p_{\rm cut}$} &
\multicolumn{1}{c}{$\langle H_0 \rangle_{\rm cut}$} &
\multicolumn{1}{c}{$\Delta\langle H_0 \rangle$} &
\multicolumn{1}{c}{$\langle V \rangle_{\rm cut}$} &
\multicolumn{1}{c}{$\Delta\langle V \rangle$} &
\multicolumn{1}{c}{$\langle H_0 \rangle_{\rm cut} + \langle V \rangle_{\rm cut}$} &
\multicolumn{1}{c}{$\Delta E$} &
\multicolumn{1}{c}{$N_P$}\\
{[fm$^{-1}$]} & {[MeV]} & {[MeV]} & {[MeV]} & {[MeV]} & {[MeV]} & {[MeV]} & \\
\hline\\[-8pt]
  50 & 13.35627892 & 1.93e-4  & -15.58065272 & -3.87e-4 & -2.22437379 & -1.93e-4  &  48\\
 100 & 13.35644813 & 2.42e-5  & -15.58099114 & -4.84e-5 & -2.22454300 & -2.42e-5  &  48\\
 200 & 13.35646930 & 3.02e-6  & -15.58103347 & -6.05e-6 & -2.22456417 & -3.02e-6  &  64\\
 500 & 13.35647213 & 1.94e-7  & -15.58103914 & -3.87e-7 & -2.22456700 & -1.94e-7  &  96\\
1000 & 13.35647230 & 2.49e-8  & -15.58103947 & -4.97e-8 & -2.22456717 & -2.49e-8  & 128\\
2000 & 13.35647232 & 3.20e-9  & -15.58103952 & -6.39e-9 & -2.22456719 & -3.20e-9  & 192\\
3000 & 13.35647233 & 9.15e-10 & -15.58103952 & -1.83e-9 & -2.22456720 & -9.15e-10 & 256\\
\end{tabular*}
\end{ruledtabular}
\\[-2pt]
\end{table*}
\begin{table*}[htbp]
\caption{
\linespread{1.3}\selectfont
Momentum-space convergence for YAMA-23 derived from the numerical solution
of the Lippmann--Schwinger Eq.~(\ref{eq:lspsinotensor}). Definitions of
$\Delta$-quantities and exact values follow
Table~\ref{tab:momentum_convergence_YAMA23}.}
\label{tab:ls_convergence_YAMA23}
\begin{ruledtabular}
\setlength{\tabcolsep}{0pt}
\sisetup{exponent-product = \cdot}
\begin{tabular*}{\textwidth}{
    @{\extracolsep{\fill}}
    S[table-format=4.0, group-digits=false]
    S[table-format=2.8, group-digits=false]
    S[table-format=1.2e-1, group-digits=false]
    S[table-format=-2.8, group-digits=false]
    S[table-format=-1.2e-1, group-digits=false]
    S[table-format=-1.8, group-digits=false]
    S[table-format=-1.2e-2, group-digits=false]
    S[table-format=3.0, group-digits=false]
}
\multicolumn{1}{c}{$p_{\rm cut}$} &
\multicolumn{1}{c}{$\langle H_0 \rangle_{\rm cut}$} &
\multicolumn{1}{c}{$\Delta\langle H_0 \rangle$} &
\multicolumn{1}{c}{$\langle V \rangle_{\rm cut}$} &
\multicolumn{1}{c}{$\Delta\langle V \rangle$} &
\multicolumn{1}{c}{$\langle H_0 \rangle_{\rm cut} + \langle V \rangle_{\rm cut}$} &
\multicolumn{1}{c}{$\Delta E$} &
\multicolumn{1}{c}{$N_P$}\\
{[fm$^{-1}$]} & {[MeV]} & {[MeV]} & {[MeV]} & {[MeV]} & {[MeV]} & {[MeV]} & \\
\hline\\[-8pt]
  50 & 13.35569835 & 7.74e-4 & -15.58007215 & -9.67e-4 & -2.22437380 & -1.93e-4  &  48\\
 100 & 13.35637551 & 9.68e-5 & -15.58091851 & -1.21e-4 & -2.22454300 & -2.42e-5  &  48\\
 200 & 13.35646022 & 1.21e-5 & -15.58102439 & -1.51e-5 & -2.22456417 & -3.02e-6  &  64\\
 500 & 13.35647155 & 7.77e-7 & -15.58103855 & -9.71e-7 & -2.22456700 & -1.94e-7  &  96\\
1000 & 13.35647223 & 9.68e-8 & -15.58103940 & -1.21e-7 & -2.22456717 & -2.42e-8  & 192\\
2000 & 13.35647231 & 1.21e-8 & -15.58103951 & -1.51e-8 & -2.22456719 & -3.03e-9  & 256\\
3000 & 13.35647232 & 3.96e-9 & -15.58103952 & -4.95e-9 & -2.22456720 & -9.91e-10 & 256\\
\end{tabular*}
\end{ruledtabular}
\\[-2pt]
\end{table*}
\begin{table*}[htbp]
\caption{
\linespread{1.3}\selectfont
Coordinate-space convergence for YAMA-23 using the analytical wave function
given in Eq.~(\ref{eq:psirspace}). Definitions of $\Delta$-quantities follow
Table~\ref{tab:momentum_convergence_YAMA23}. Exact $r_{\rm cut}\to\infty$
limits are given in Appendix~\ref{sec:waverspace}.}
\label{tab:rspace_convergence_YAMA23}
\begin{ruledtabular}
\setlength{\tabcolsep}{0pt}
\sisetup{exponent-product = \cdot}
\begin{tabular*}{\textwidth}{
    @{\extracolsep{\fill}}
    S[table-format=2.0, group-digits=false]
    S[table-format=2.8, group-digits=false]
    S[table-format=-1.2e-2, group-digits=false]
    S[table-format=-2.8, group-digits=false]
    S[table-format=-1.2e-2, group-digits=false]
    S[table-format=-1.8, group-digits=false]
    S[table-format=-1.2e-2, group-digits=false]
    S[table-format=2.0, group-digits=false]
}
\multicolumn{1}{c}{$r_{\rm cut}$} &
\multicolumn{1}{c}{$\langle H_0 \rangle_{\rm cut}$} &
\multicolumn{1}{c}{$\Delta\langle H_0 \rangle$} &
\multicolumn{1}{c}{$\langle V \rangle_{\rm cut}$} &
\multicolumn{1}{c}{$\Delta\langle V \rangle$} &
\multicolumn{1}{c}{$\langle H_0 \rangle_{\rm cut} + \langle V \rangle_{\rm cut}$} &
\multicolumn{1}{c}{$\Delta E$} &
\multicolumn{1}{c}{$N_R$} \\
{[fm]} & {[MeV]} & {[MeV]} & {[MeV]} & {[MeV]} & {[MeV]} & {[MeV]} & \\
\hline\\[-8pt]
10 & 13.39283416 & -3.64e-2  & -15.58103278 & -6.74e-6  & -2.18819863 & -3.64e-2  & 32 \\
20 & 13.35682631 & -3.54e-4  & -15.58103952 & -6.09e-13 & -2.22421321 & -3.54e-4  & 32 \\
30 & 13.35647577 & -3.45e-6  & -15.58103952 & -1.78e-15 & -2.22456375 & -3.45e-6  & 32 \\
40 & 13.35647236 & -3.35e-8  & -15.58103952 & -1.78e-15 & -2.22456716 & -3.35e-8  & 32 \\
50 & 13.35647233 & -3.26e-10 & -15.58103952 & -1.78e-15 & -2.22456720 & -3.26e-10 & 48 \\
60 & 13.35647233 & -5.87e-12 & -15.58103952 &  5.24e-12 & -2.22456720 & -6.30e-13 & 48\\
\end{tabular*}
\end{ruledtabular}
\\[-2pt]
\end{table*}
\begin{table*}[htbp]
\caption{
\linespread{1.3}\selectfont
Verification of the numerical Fourier transform for YAMA-23. The numerical
solution of the Lippmann--Schwinger Eq.~(\ref{eq:lspsinotensor}) is
transformed to coordinate space, and integrals are evaluated on $[0,r_{\rm cut}]$.
Definitions of $\Delta$-quantities and exact values follow
Table~\ref{tab:momentum_convergence_YAMA23}.}
\label{tab:rspace_ls_convergence_yama23}
\begin{ruledtabular}
\setlength{\tabcolsep}{0pt}
\sisetup{exponent-product = \cdot}
\begin{tabular*}{\textwidth}{
    @{\extracolsep{\fill}}
    S[table-format=2.0, group-digits=false]
    S[table-format=2.8, group-digits=false]
    S[table-format=-1.2e-1, group-digits=false]
    S[table-format=-2.8, group-digits=false]
    S[table-format=-1.2e-1, group-digits=false]
    S[table-format=-1.8, group-digits=false]
    S[table-format=-1.2e-1, group-digits=false]
    S[table-format=2.0, group-digits=false]
}
\multicolumn{1}{c}{$r_{\rm cut}$} &
\multicolumn{1}{c}{$\langle H_0 \rangle_{\rm cut}$} &
\multicolumn{1}{c}{$\Delta\langle H_0 \rangle$} &
\multicolumn{1}{c}{$\langle V \rangle_{\rm cut}$} &
\multicolumn{1}{c}{$\Delta\langle V \rangle$} &
\multicolumn{1}{c}{$\langle H_0 \rangle_{\rm cut} + \langle V \rangle_{\rm cut}$} &
\multicolumn{1}{c}{$\Delta E$} &
\multicolumn{1}{c}{$N_R$}\\
{[fm]} & {[MeV]} & {[MeV]} & {[MeV]} & {[MeV]} & {[MeV]} & {[MeV]} & \\
\hline\\[-8pt]
10 & 13.39420424 & -3.77e-2 & -15.58103278 & -6.75e-6 & -2.18682854 & -3.77e-2 & 32\\
20 & 13.35685907 & -3.87e-4 & -15.58103952 & -6.89e-9 & -2.22418045 & -3.87e-4 & 32\\
30 & 13.35647612 & -3.79e-6 & -15.58103952 & -6.73e-9 & -2.22456340 & -3.80e-6 & 32\\
40 & 13.35647241 & -8.43e-8 & -15.58103952 & -7.38e-9 & -2.22456711 & -9.17e-8 & 48\\
50 & 13.35647238 & -5.81e-8 & -15.58103952 & -7.38e-9 & -2.22456713 & -6.55e-8 & 48\\
60 & 13.35647238 & -4.97e-8 & -15.58103952 & -7.26e-9 & -2.22456714 & -5.70e-8 & 48\\
\end{tabular*}
\end{ruledtabular}
\\[-2pt]
\end{table*}

Before solving the full integral equation, it is useful to quantify the
systematic error introduced by truncating the momentum integration at a finite
cut-off $p_{\rm cut}$. For the Yamaguchi potential,
Appendix~\ref{sec:limityamaguchi} provides closed-form expressions for the
cut-off dependence of $\langle H_0\rangle_{\rm cut}$ and $\langle V\rangle_{\rm
cut}$, and thus for the missing high-momentum contributions. In practice, the
same cut-off quantities can be obtained by numerically integrating the
analytical wave function, Eq.~(\ref{eq:psiyamaguchipspace}), provided the
momentum mesh is sufficiently dense. Throughout this section, we use the
convention $\Delta X \equiv X - X_{\rm cut}$, where $X$ denotes the exact limit
$p_{\rm cut}\to\infty$.

Table~\ref{tab:momentum_convergence_YAMA23} summarizes the resulting cut-off
dependence for the YAMA-23 potential and also documents the number of
Gauss--Legendre mesh points $N$ required to converge the numerical integrals for
a given $p_{\rm cut}$. For example, $p_{\rm cut}=50$~fm$^{-1}$ yields an energy
error of order $10^{-4}$~MeV, while $p_{\rm cut}=200$~fm$^{-1}$ reduces the
error to the $10^{-6}$~MeV level. The table also illustrates the interplay
between cut-off and resolution. While $N=48$ is sufficient to converge the
integrals for $p_{\rm cut}=100$~fm$^{-1}$, substantially larger meshes are
required when the integration range is extended to very large cut-offs.

The crucial test is the numerical solution of the discretized
Lippmann--Schwinger equation~(\ref{eq:lspsinotensor}).
Table~\ref{tab:ls_convergence_YAMA23} shows the expectation values obtained
from the eigenvalue calculation over the same set of cut-offs. A direct
comparison with Table~\ref{tab:momentum_convergence_YAMA23} shows that, for
each $p_{\rm cut}$, the converged energy sum $\langle H_0\rangle_{\rm
cut}+\langle V\rangle_{\rm cut}$ reproduces the corresponding cut-off benchmark.
In particular, the deviation $\Delta E_2$ agrees within the tabulated precision.
This indicates that the discretized eigenvalue solution introduces no additional
significant numerical error beyond the truncation at $p_{\rm cut}$.

As a final check, we analyze the bound-state wave function in coordinate space.
The coordinate-space wave function $\psi(r)$ is obtained by numerically Fourier
transforming the momentum-space solution $\psi(p)$ using
Eq.~(\ref{eq:psifourier}). For the Yamaguchi potential this transformation can
also be carried out analytically, providing an additional exact benchmark.
Moreover, Appendix~\ref{sec:waverspace} provides closed-form expressions for
the $r_{\rm cut}$ dependence of the relevant expectation values, allowing us to
separate truncation effects from discretization errors.

We first validate the coordinate-space quadrature and the sole effect of
truncating the integration at $r_{\rm cut}$ by integrating the analytical
wave function $\psi(r)$ (Eq.~\ref{eq:psirspace}) directly. For the
numerical evaluation of observables in coordinate space, the integration
over the radial variable $r$ is discretized using a Gauss--Legendre
mesh with $N_R$ points on the interval $[0, r_{\rm cut}]$.
Table~\ref{tab:rspace_convergence_YAMA23} lists the cut-off expectation values
and their deviations from the exact $r_{\rm cut}\to\infty$ limits. It shows,
for example, that $r_{\rm cut}\approx 20$~fm is required to reduce the energy
error to the $10^{-4}$~MeV level, while $30$~fm reduces it to $10^{-6}$~MeV.
The table also documents the minimum number of Gauss--Legendre points $N_R$
needed to converge the integrals for each cut-off. This provides a baseline
accuracy that a numerical Fourier transformation and subsequent
integrations cannot exceed.

\begin{table*}[htbp]
\caption{
\linespread{1.3}\selectfont
Coordinate-space convergence of radius observables for YAMA-23 using the
analytical wave function from Eq.~(\ref{eq:psirspace}) for the numerical
integration. $\Delta$-quantities are relative to the exact limits from
Eqs.~(\ref{eq:limitradius})--(\ref{eq:limitrmsradius}). The exact values are
$\langle \tfrac{1}{2}r\rangle = 1.56742466$~fm and
$\sqrt{\langle (\tfrac{1}{2}r)^2\rangle} = 1.93630777$~fm. $N_R$ denotes the
number of radial mesh points in coordinate space.}
\label{tab:rspace_convergence_yama23_radius}
\begin{ruledtabular}
\setlength{\tabcolsep}{0pt}
\sisetup{exponent-product = \cdot}
\begin{tabular*}{\textwidth}{
    @{\extracolsep{\fill}}
    S[table-format=2.0, group-digits=false]
    S[table-format=1.8, group-digits=false]
    S[table-format=1.2e-2, group-digits=false]
    S[table-format=1.8, group-digits=false]
    S[table-format=1.2e-1, group-digits=false]
    S[table-format=2.0, group-digits=false]
}
\multicolumn{1}{c}{$r_{\rm cut}$} &
\multicolumn{1}{c}{$\langle \tfrac{1}{2}r \rangle_{\rm cut}$} &
\multicolumn{1}{c}{$\Delta\langle \tfrac{1}{2}r \rangle$} &
\multicolumn{1}{c}{$\sqrt{\langle (\tfrac{1}{2}r)^2 \rangle_{\rm cut}}$} &
\multicolumn{1}{c}{$\Delta\sqrt{\langle (\tfrac{1}{2}r)^2 \rangle}$} &
\multicolumn{1}{c}{$N_R$}\\
{[fm]} & {[fm]} & {[fm]} & {[fm]} & {[fm]} & \\
\hline\\[-8pt]
10 & 1.46804380 & 9.94e-2  & 1.76806754 & 1.68e-1 & 32\\
20 & 1.56566166 & 1.76e-3  & 1.93120929 & 5.10e-3 & 32\\
30 & 1.56739976 & 2.49e-5  & 1.93620389 & 1.04e-4 & 32\\
40 & 1.56742434 & 3.18e-7  & 1.93630604 & 1.73e-6 & 32\\
50 & 1.56742466 & 4.09e-9  & 1.93630775 & 2.60e-8 & 32\\
60 & 1.56742466 & 2.33e-10 & 1.93630777 & 2.35e-9 & 32\\
\end{tabular*}
\end{ruledtabular}
\\[-2pt]
\end{table*}

\begin{table*}[htbp]
\caption{
\linespread{1.3}\selectfont
Coordinate-space convergence of radius observables for the YAMA-23 potential,
obtained by Fourier transforming the numerical solution of the
Lippmann--Schwinger Eq.~(\ref{eq:lspsinotensor}). Definitions of
$\Delta$-quantities and exact values follow
Table~\ref{tab:rspace_convergence_yama23_radius}. $N_R$ denotes the
number of radial mesh points in coordinate space.}
\label{tab:fourier_LS_rspace_convergence_yama23_radius}
\begin{ruledtabular}
\setlength{\tabcolsep}{0pt}
\sisetup{exponent-product = \cdot}
\begin{tabular*}{\textwidth}{
    @{\extracolsep{\fill}}
    S[table-format=2.0, group-digits=false]
    S[table-format=1.8, group-digits=false]
    S[table-format=1.2e-2, group-digits=false]
    S[table-format=1.8, group-digits=false]
    S[table-format=1.2e-2, group-digits=false]
    S[table-format=2.0, group-digits=false]
}
\multicolumn{1}{c}{$r_{\rm cut}$} &
\multicolumn{1}{c}{$\langle \tfrac{1}{2}r \rangle_{\rm cut}$} &
\multicolumn{1}{c}{$\Delta\langle \tfrac{1}{2}r \rangle$} &
\multicolumn{1}{c}{$\sqrt{\langle (\tfrac{1}{2}r)^2 \rangle_{\rm cut}}$} &
\multicolumn{1}{c}{$\Delta\sqrt{\langle (\tfrac{1}{2}r)^2 \rangle}$} &
\multicolumn{1}{c}{$N_R$}\\
{[fm]} & {[fm]} & {[fm]} & {[fm]} & {[fm]} & \\
\hline\\[-8pt]
10 & 1.46804380 & 9.94e-2  & 1.76806754 & 1.68e-1  & 32\\
20 & 1.56566166 & 1.76e-3  & 1.93120929 & 5.10e-3  & 32\\
30 & 1.56739976 & 2.49e-5  & 1.93620389 & 1.04e-4  & 32\\
40 & 1.56742434 & 3.18e-7  & 1.93630604 & 1.73e-6  & 32\\
50 & 1.56742466 & 3.86e-9  & 1.93630775 & 2.58e-8  & 48\\
60 & 1.56742466 & 4.68e-11 & 1.93630777 & 3.01e-10 & 48\\
\end{tabular*}
\end{ruledtabular}
\\[-2pt]
\end{table*}

\begin{table*}[htbp]
\caption{
\linespread{1.3}\selectfont
Two-body binding energies $E_2$ for Yamaguchi and Malfliet--Tjon-type
potentials. 1D refers to the numerical solution of the partial-wave-decomposed
Lippmann--Schwinger Eq.~(\ref{eq:lspsinotensor}). 2D refers to the solution
of the two-dimensional Lippmann--Schwinger Eq.~(\ref{eq:ls2d}), with the
potential restricted to $l=0$, $l=12$, or without an explicit partial-wave
decomposition ($l=\infty$). For the Yamaguchi potentials, the $l=\infty$
case is not shown because the interaction is implemented here only through
its partial-wave form factors. For the Malfliet--Tjon potentials, the
momentum cut-offs are $p_{\rm cut} = 3000$~fm$^{-1}$ for MT-III and
$p_{\rm cut} = 4000$~fm$^{-1}$ for MT-IV.}
\label{tab:yama_mt2d3d}
\begin{ruledtabular}
\setlength{\tabcolsep}{0pt}
\begin{tabular*}{\textwidth}{
    @{\extracolsep{\fill}}
    l
    S[table-format=-1.12, group-digits=false]
    S[table-format=-1.10, group-digits=false]
    S[table-format=-1.10, group-digits=false]
    S[table-format=-1.10, group-digits=false]
}
Potential &
\multicolumn{1}{c}{1D} &
\multicolumn{1}{c}{2D ($l=0$)} &
\multicolumn{1}{c}{2D ($l=12$)} &
\multicolumn{1}{c}{2D ($l=\infty$)} \\
 &
\multicolumn{1}{c}{$E_2$ [MeV]} &
\multicolumn{1}{c}{$E_2$ [MeV]} &
\multicolumn{1}{c}{$E_2$ [MeV]} &
\multicolumn{1}{c}{$E_2$ [MeV]} \\
\hline\\[-8pt]
YAMA-23  &  -2.2245671961 & -2.2245671961 & -2.2245671961 & {--} \\
YAMA-IV  &  -0.3317813279 & -0.3317813279 & -0.3317813279 & {--} \\
\hline\\[-8pt]
MT-III     &  -2.4086354317 &  -2.4086354317 &  -2.4086354317 &   -2.4086354329 \\
MT-IV      &  -2.2086291832 &  -2.2086291832 &  -2.2086291832 &   -2.2086291862 \\
\end{tabular*}
\end{ruledtabular}
\\[-2pt]
\end{table*}

\begin{table*}[htbp]
\caption{
\linespread{1.3}\selectfont
Convergence of expectation values and binding energy for the MT-IV potential
with respect to $N_P$ and $N_X$ at a fixed cut-off $p_{\rm cut} = 4000$
fm$^{-1}$. The 1D partial-wave benchmark is $E_2 = -2.2086291832$~MeV.}
\label{tab:convergence_nx_np}
\begin{ruledtabular}
\setlength{\tabcolsep}{0pt}
\sisetup{exponent-product = \cdot}
\begin{tabular*}{\textwidth}{
    @{\extracolsep{\fill}}
    S[table-format=4.0]
    S[table-format=3.0]
    S[table-format=2.12, group-digits=false]
    S[table-format=-2.12, group-digits=false]
    S[table-format=-1.12, group-digits=false]
    S[table-format=-1.12, group-digits=false]
    S[table-format=1.2e-2, group-digits=false]
}
\multicolumn{1}{c}{$N_P$} & \multicolumn{1}{c}{$N_X$} &
\multicolumn{1}{c}{$\langle H_0 \rangle$} & \multicolumn{1}{c}{$\langle V \rangle$} &
\multicolumn{1}{c}{$H_0+V$} & \multicolumn{1}{c}{$E_2$} &
\multicolumn{1}{c}{$\Delta E$} \\
 & & \multicolumn{1}{c}{[MeV]} & \multicolumn{1}{c}{[MeV]} &
\multicolumn{1}{c}{[MeV]} & \multicolumn{1}{c}{[MeV]} &
\multicolumn{1}{c}{[MeV]} \\
\hline\\[-8pt]
 512 &  32 & 11.797304745964 & -14.005952579887 & -2.208647833923 & -2.208647834069 & 1.46e-10 \\
 512 &  96 & 11.797224070428 & -14.005860191647 & -2.208636121218 & -2.208636122524 & 1.31e-09 \\
1024 &  64 & 11.797178427749 & -14.005807946616 & -2.208629518867 & -2.208629519199 & 3.32e-10 \\
2048 &  64 & 11.797177664980 & -14.005807072648 & -2.208629407668 & -2.208629409030 & 1.36e-09 \\
2048 &  96 & 11.797176308454 & -14.005805518563 & -2.208629210109 & -2.208629212894 & 2.78e-09 \\
2048 & 128 & 11.797176173381 & -14.005805362681 & -2.208629189300 & -2.208629193387 & 4.09e-09 \\
2048 & 160 & 11.797176123322 & -14.005805305771 & -2.208629182449 & -2.208629186158 & 3.71e-09 \\
\end{tabular*}
\end{ruledtabular}
\\[-2pt]
\end{table*}

\begin{table*}[htbp]
\caption{
\linespread{1.3}\selectfont
Momentum-space convergence for the MT-III potential derived from the numerical
solution of the Lippmann--Schwinger Eq.~(\ref{eq:lspsinotensor}). Expectation
values are evaluated on $[0,p_{\rm cut}]$ using $N_P$ mesh points.
$\Delta E \equiv E_2 - (\langle H_0\rangle_{\rm cut}+\langle V\rangle_{\rm cut})$
quantifies the internal consistency.}
\label{tab:ls_convergence_MT3}
\begin{ruledtabular}
\setlength{\tabcolsep}{0pt}
\sisetup{exponent-product = \cdot}
\begin{tabular*}{\textwidth}{
    @{\extracolsep{\fill}}
    S[table-format=4.0, group-digits=false]
    S[table-format=2.8, group-digits=false]
    S[table-format=-2.8, group-digits=false]
    S[table-format=-1.8, group-digits=false]
    S[table-format=-1.8, group-digits=false]
    S[table-format=-1.2e-2, group-digits=false]
    S[table-format=4.0, group-digits=false]
}
\multicolumn{1}{c}{$p_{\rm cut}$} &
\multicolumn{1}{c}{$\langle H_0 \rangle_{\rm cut}$} &
\multicolumn{1}{c}{$\langle V \rangle_{\rm cut}$} &
\multicolumn{1}{c}{$\langle H_0 \rangle_{\rm cut} + \langle V \rangle_{\rm cut}$} &
\multicolumn{1}{c}{$E_2$} &
\multicolumn{1}{c}{$\Delta E$} &
\multicolumn{1}{c}{$N_P$}\\
{[fm$^{-1}$]} & {[MeV]} & {[MeV]} & {[MeV]} & {[MeV]} & {[MeV]} & \\
\hline\\[-8pt]
  50 & 11.22961108 & -13.63814854 & -2.40853746 & -2.40853746 &  2.75e-14 &  128\\
 100 & 11.22965900 & -13.63828308 & -2.40862408 & -2.40862408 &  2.49e-14 &  256\\
 500 & 11.22966464 & -13.63829999 & -2.40863535 & -2.40863535 &  7.99e-15 &  768\\
1000 & 11.22966468 & -13.63830010 & -2.40863542 & -2.40863542 &  2.71e-14 &  768\\
1500 & 11.22966468 & -13.63830011 & -2.40863543 & -2.40863543 & -1.33e-14 &  768\\
2000 & 11.22966468 & -13.63830011 & -2.40863543 & -2.40863543 & -1.33e-14 & 1024\\
\end{tabular*}
\end{ruledtabular}
\\[-2pt]
\end{table*}

\begin{table*}[htbp]
\caption{
\linespread{1.3}\selectfont
Same as Table~\ref{tab:ls_convergence_MT3}, but for the MT-IV potential.\hfill}
\label{tab:ls_convergence_MT4}
\begin{ruledtabular}
\setlength{\tabcolsep}{0pt}
\sisetup{exponent-product = \cdot}
\begin{tabular*}{\textwidth}{
    @{\extracolsep{\fill}}
    S[table-format=4.0, group-digits=false]
    S[table-format=2.8, group-digits=false]
    S[table-format=-2.8, group-digits=false]
    S[table-format=-1.8, group-digits=false]
    S[table-format=-1.8, group-digits=false]
    S[table-format=-1.2e-2, group-digits=false]
    S[table-format=4.0, group-digits=false]
}
\multicolumn{1}{c}{$p_{\rm cut}$} &
\multicolumn{1}{c}{$\langle H_0 \rangle_{\rm cut}$} &
\multicolumn{1}{c}{$\langle V \rangle_{\rm cut}$} &
\multicolumn{1}{c}{$\langle H_0 \rangle_{\rm cut} + \langle V \rangle_{\rm cut}$} &
\multicolumn{1}{c}{$E_2$} &
\multicolumn{1}{c}{$\Delta E$} &
\multicolumn{1}{c}{$N_P$}\\
{[fm$^{-1}$]} & {[MeV]} & {[MeV]} & {[MeV]} & {[MeV]} & {[MeV]} & \\
\hline\\[-8pt]
  50 & 11.79629175 & -14.00477146 & -2.20847971 & -2.20847971 &  <1.0e-16 &  256\\
 100 & 11.79706447 & -14.00567481 & -2.20861034 & -2.20861034 & -2.66e-15 &  512\\
 500 & 11.79717521 & -14.00580424 & -2.20862903 & -2.20862903 & -2.13e-14 &  768\\
1000 & 11.79717599 & -14.00580516 & -2.20862916 & -2.20862916 & -1.64e-14 & 1536\\
1500 & 11.79717607 & -14.00580525 & -2.20862918 & -2.20862918 & -3.15e-14 & 1536\\
2000 & 11.79717609 & -14.00580527 & -2.20862918 & -2.20862918 &  8.44e-15 & 2048\\
3000 & 11.79717610 & -14.00580529 & -2.20862918 & -2.20862918 &  4.80e-14 & 2048\\
3500 & 11.79717611 & -14.00580529 & -2.20862918 & -2.20862918 & -4.35e-14 & 2048\\
4000 & 11.79717611 & -14.00580529 & -2.20862918 & -2.20862918 & -1.15e-14 & 2048\\
\end{tabular*}
\end{ruledtabular}
\\[-2pt]
\end{table*}

\begin{table*}[htbp]
\caption{
\linespread{1.3}\selectfont
Coordinate-space convergence for the MT-III potential, derived from the
Fourier-transformed numerical solution of the Lippmann--Schwinger
Eq.~(\ref{eq:lspsinotensor}). Expectation values are evaluated on
$[0,r_{\rm cut}]$ using $N_R$ mesh points. $\Delta E$ quantifies the consistency
with the momentum-space binding energy $E_2$.}
\label{tab:fourier_rspace_convergence_mt3}
\begin{ruledtabular}
\setlength{\tabcolsep}{0pt}
\sisetup{exponent-product = \cdot}
\begin{tabular*}{\textwidth}{
    @{\extracolsep{\fill}}
    S[table-format=2.0, group-digits=false]
    S[table-format=2.8, group-digits=false]
    S[table-format=-2.8, group-digits=false]
    S[table-format=-1.8, group-digits=false]
    S[table-format=-1.8, group-digits=false]
    S[table-format=-1.2e-1, group-digits=false]
    S[table-format=2.0, group-digits=false]
}
\multicolumn{1}{c}{$r_{\rm cut}$} &
\multicolumn{1}{c}{$\langle H_0 \rangle_{\rm cut}$} &
\multicolumn{1}{c}{$\langle V \rangle_{\rm cut}$} &
\multicolumn{1}{c}{$\langle H_0 \rangle_{\rm cut} + \langle V \rangle_{\rm cut}$} &
\multicolumn{1}{c}{$E_2$} &
\multicolumn{1}{c}{$\Delta E$} &
\multicolumn{1}{c}{$N_R$}\\
{[fm]} & {[MeV]} & {[MeV]} & {[MeV]} & {[MeV]} & {[MeV]} & \\
\hline\\[-8pt]
10 & 11.26603326 & -13.63830007 & -2.37226681 & -2.40863543 & -3.64e-2 & 32\\
20 & 11.22997413 & -13.63830011 & -2.40832597 & -2.40863543 & -3.09e-4 & 32\\
30 & 11.22966715 & -13.63830011 & -2.40863295 & -2.40863543 & -2.48e-6 & 32\\
40 & 11.22966474 & -13.63830011 & -2.40863537 & -2.40863543 & -6.07e-8 & 48\\
50 & 11.22966471 & -13.63830011 & -2.40863540 & -2.40863543 & -3.41e-8 & 48\\
60 & 11.22966471 & -13.63830011 & -2.40863539 & -2.40863543 & -4.02e-8 & 48\\
\end{tabular*}
\end{ruledtabular}
\\[-2pt]
\end{table*}

\begin{table*}[htbp]
\caption{
\linespread{1.3}\selectfont
Same as Table~\ref{tab:fourier_rspace_convergence_mt3}, but for the MT-IV potential.\hfill}
\label{tab:fourier_rspace_convergence_mt4}
\begin{ruledtabular}
\setlength{\tabcolsep}{0pt}
\sisetup{exponent-product = \cdot}
\begin{tabular*}{\textwidth}{
    @{\extracolsep{\fill}}
    S[table-format=2.0, group-digits=false]
    S[table-format=2.8, group-digits=false]
    S[table-format=-2.8, group-digits=false]
    S[table-format=-1.8, group-digits=false]
    S[table-format=-1.8, group-digits=false]
    S[table-format=-1.2e-1, group-digits=false]
    S[table-format=2.0, group-digits=false]
}
\multicolumn{1}{c}{$r_{\rm cut}$} &
\multicolumn{1}{c}{$\langle H_0 \rangle_{\rm cut}$} &
\multicolumn{1}{c}{$\langle V \rangle_{\rm cut}$} &
\multicolumn{1}{c}{$\langle H_0 \rangle_{\rm cut} + \langle V \rangle_{\rm cut}$} &
\multicolumn{1}{c}{$E_2$} &
\multicolumn{1}{c}{$\Delta E$} &
\multicolumn{1}{c}{$N_R$}\\
{[fm]} & {[MeV]} & {[MeV]} & {[MeV]} & {[MeV]} & {[MeV]} & \\
\hline\\[-8pt]
10 & 11.83909961 & -14.00572261 & -2.16662300 & -2.20862918 & -4.20e-2 & 32\\
20 & 11.79761395 & -14.00580528 & -2.20819133 & -2.20862918 & -4.38e-4 & 32\\
30 & 11.79718040 & -14.00580528 & -2.20862488 & -2.20862918 & -4.30e-6 & 48\\
40 & 11.79717619 & -14.00580528 & -2.20862909 & -2.20862918 & -8.90e-8 & 48\\
50 & 11.79717616 & -14.00580528 & -2.20862913 & -2.20862918 & -5.60e-8 & 48\\
60 & 11.79717615 & -14.00580528 & -2.20862914 & -2.20862918 & -4.80e-8 & 48\\
\end{tabular*}
\end{ruledtabular}
\\[-2pt]
\end{table*}

\begin{table*}[htbp]
\caption{
\linespread{1.3}\selectfont
Coordinate-space convergence of radius observables for the MT-III and MT-IV
potentials, obtained by Fourier transforming the numerical solution of the
Lippmann--Schwinger Eq.~(\ref{eq:lspsinotensor}). The radius
$\langle \tfrac{1}{2}r\rangle_{\rm cut}$ and rms radius
$\sqrt{\langle(\tfrac{1}{2}r)^2 \rangle_{\rm cut}}$ are evaluated on
$[0,r_{\rm cut}]$ using $N_R$ mesh points.}
\label{tab:fourier_rspace_convergence_mt34_radius}
\begin{ruledtabular}
\setlength{\tabcolsep}{0pt}
\begin{tabular*}{\textwidth}{
    @{\extracolsep{\fill}}
    S[table-format=2.0, group-digits=false]
    S[table-format=1.8, group-digits=false]
    S[table-format=1.8, group-digits=false]
    S[table-format=2.0, group-digits=false]
    S[table-format=1.8, group-digits=false]
    S[table-format=1.8, group-digits=false]
    S[table-format=2.0, group-digits=false]
}
 & \multicolumn{3}{c}{MT-III} & \multicolumn{3}{c}{MT-IV} \\
\cline{2-4} \cline{5-7}\\[-8pt]
\multicolumn{1}{c}{$r_{\rm cut}$} &
\multicolumn{1}{c}{$\langle \tfrac{1}{2}r \rangle_{\rm cut}$} &
\multicolumn{1}{c}{$\sqrt{\langle (\tfrac{1}{2}r)^2 \rangle_{\rm cut}}$} &
\multicolumn{1}{c}{$N_R$} &
\multicolumn{1}{c}{$\langle \tfrac{1}{2}r \rangle_{\rm cut}$} &
\multicolumn{1}{c}{$\sqrt{\langle (\tfrac{1}{2}r)^2 \rangle_{\rm cut}}$} &
\multicolumn{1}{c}{$N_R$}\\
{[fm]} & {[fm]} & {[fm]} & & {[fm]} & {[fm]} & \\
\hline\\[-8pt]
10 & 1.51978984 & 1.78605363 & 32 &  1.53035364 & 1.83041195 & 32\\
20 & 1.60626657 & 1.92902125 & 32 &  1.63986398 & 2.00695201 & 32\\
30 & 1.60754579 & 1.93269021 & 32 &  1.64184639 & 2.01243497 & 32\\
40 & 1.60756080 & 1.93275254 & 32 &  1.64187490 & 2.01254894 & 32\\
50 & 1.60756096 & 1.93275340 & 32 &  1.64187527 & 2.01255088 & 32\\
60 & 1.60756096 & 1.93275341 & 48 &  1.64187528 & 2.01255091 & 48\\
70 & 1.60756096 & 1.93275341 & 48 &  1.64187528 & 2.01255091 & 48\\
\end{tabular*}
\end{ruledtabular}
\\[-2pt]
\end{table*}

Next, we isolate numerical effects originating from the Fourier transformation
itself. To this end, we Fourier transform the exact analytical momentum-space
wave function $\psi(p)$ from Eq.~(\ref{eq:psiyamaguchipspace}) to coordinate
space. The radial Fourier integrals for the reduced wave function $u(r)$ and
its derivative $u'(r)$ are evaluated using Simpson quadrature in the
dimensionless variable $x = p \, r$. This ensures that the oscillations
of the spherical Bessel functions are resolved with a constant number of
grid points per period for any distance $r$.

For the evaluation of the kinetic-energy expectation value, the operator is
rewritten in a form involving only first derivatives. To avoid numerical
differentiation in coordinate space, $u'(r)$ is calculated directly from the
Fourier transform by using the analytical derivative of the spherical Bessel
function inside the Fourier integral. When applying this to numerical
solutions, $\psi(p)$ is interpolated by cubic splines. This method avoids the
amplification of interpolation errors and ensures that the coordinate-space
expectation values are fully consistent with the momentum-space results.

The coordinate-space observables obtained by transforming the analytical wave
function were found to agree within numerical precision to those derived from
the numerical Lippmann--Schwinger solution. This confirms that the observed
deviations are dominated by the Fourier transformation and coordinate-space
integration rather than the momentum-space discretization. Consequently, we
present the results for the full numerical procedure in
Table~\ref{tab:rspace_ls_convergence_yama23}. Comparing these results with
the analytical baseline established in
Table~\ref{tab:rspace_convergence_YAMA23} reveals that the remaining
deviations are consistent with the truncation at $r_{\rm cut}$. The high
precision confirms that the use of the exact Fourier derivative eliminates the
numerical noise typically associated with kinetic-energy evaluation in
coordinate space.

Complementing the energy analysis, we calculate the mean radius $\langle
\tfrac{1}{2}r\rangle$ and the rms radius $\sqrt{\langle
(\tfrac{1}{2}r)^2\rangle}$ in coordinate space. To validate these observables,
we employ a multi-stage strategy analogous to the energy analysis. First, we
establish the exact benchmark values from the analytical $r_{\rm cut}\to\infty$
limits derived in Appendix~\ref{sec:waverspace} (Eqs.~\ref{eq:limitradius} and
\ref{eq:limitrmsradius}). Second, we validate the coordinate-space quadrature
by numerically integrating the analytical wave function $\psi(r)$,
Eq.~(\ref{eq:psirspace}), up to $r_{\rm cut}$. The results are shown in
Table~\ref{tab:rspace_convergence_yama23_radius} and demonstrate rapid
convergence with increasing cut-off.

Next, we compute the radii by Fourier transforming both the analytical and
the numerically determined momentum-space wave functions. As with the energies,
these two methods yield results that agree within numerical precision,
confirming the accuracy of the transformation.
Table~\ref{tab:fourier_LS_rspace_convergence_yama23_radius}
shows the radii obtained from the full numerical solution. The values agree
with the analytical benchmarks within the expected finite-$r_{\rm cut}$
truncation error. This confirms that the momentum-space solution, the Fourier
transformation, and the coordinate-space integrations yield consistent radii.

Finally, Table~\ref{tab:yama_mt2d3d} demonstrates the numerical
equivalence of the one- and two-dimensional formulations for the Yamaguchi
potential. Since the interaction acts purely in the $s$-wave, the binding
energies obtained from the 2D equation, both restricted to $l=0$ and with higher
partial waves included up to $l=12$, match the 1D partial-wave result to at
least ten significant digits. This confirms that the vector-variable formulation
correctly recovers the partial-wave limit and that the angular integration grid
is sufficiently dense. The remaining deviation from the exact analytical value
in the tenth decimal place is common to both formulations and arises solely from
the discretization of the radial momentum integral.

To further verify the internal consistency of the 2D approach, we analyzed the
angular content of the resulting wave function $\psi(p,x)$. Inverting the expansion
in Eq.~(\ref{eq:psi2dpartial}), the partial-wave components are given by
\begin{equation}
\psi_l(p) = \sqrt{\pi \, (2  l + 1)}\int\limits_{-1}^{1} \! dx \, \psi(p,x) \, P_l(x) \,.
\end{equation}
Numerical evaluation of this integral confirms that the contributions of higher
partial waves ($l>0$) to the full wave function are suppressed to the level of
machine precision ($< 10^{-30}$). This strict preservation of the $s$-wave
character within the full vector formalism confirms the numerical consistency
of the implementation.

\subsection{Application to the Malfliet--Tjon Potential}

Having validated our numerical framework against the analytically solvable
Yamaguchi potential, we now apply it to the local Malfliet--Tjon (MT)
interactions. In contrast to the separable Yamaguchi form factors, the MT
potentials behave as $1/r$ at short distances. In momentum space this leads to
Yukawa-type matrix elements that decay only as $\sim (p^2+\mu^2)^{-1}$ at large
momenta. As a consequence, both the convergence with respect to the momentum
cut-off $p_{\rm cut}$ and the required quadrature resolution are expected to be
significantly more demanding than for the Yamaguchi benchmarks.

We solve the 1D partial-wave Lippmann--Schwinger
equation~(\ref{eq:lspsinotensor}) for the MT parameter sets listed in
Table~\ref{tab:mtparams}. The converged binding energies as a function of
$p_{\rm cut}$ are summarized in Tables~\ref{tab:ls_convergence_MT3} and
\ref{tab:ls_convergence_MT4}. Two observations are central. First, the MT
binding energies show a noticeably stronger dependence on $p_{\rm cut}$ than the
Yamaguchi results. Whereas YAMA-23 reaches $10^{-6}$~MeV stability already at
$p_{\rm cut}\simeq 200$~fm$^{-1}$, the MT potentials require cut-offs in the
several-hundred to $\mathcal{O}(10^3)$~fm$^{-1}$ range before the corresponding
changes become negligible at the $10^{-5}$--$10^{-7}$~MeV level. This behavior
reflects the harder, local interaction.

Second, the numerical cost is dominated by the mesh resolution required to
represent the kernel accurately over the enlarged momentum interval. As seen in
Tables~\ref{tab:ls_convergence_MT3} and \ref{tab:ls_convergence_MT4},
increasing $p_{\rm cut}$ necessitates very large numbers of Gauss--Legendre
points $N_P$ to maintain the same level of convergence. To resolve the short-range
physics of the MT potentials, we found $p_0 = 9.0$~fm$^{-1}$ to be an optimal
choice for the mesh partition. For the high-precision convergence tests, a dense
momentum grid of $N_P = 2048$ points was employed, with $255$ points distributed
in $(0, p_0)$ and $1792$ points covering the high-momentum tail
$(p_0, p_{\rm cut})$, plus the additional point at the origin. Including the
point at the origin, this provides a total of 2048 mesh
points. Thus, for MT interactions, the cut-off must be chosen together with a
sufficiently dense discretization; otherwise the apparent stability in $E_2$
can be misleading.

To document the convergence of the two-dimensional calculation with respect to
the momentum and angular meshes, Table~\ref{tab:convergence_nx_np} collects
representative results for the MT-IV potential at fixed
$p_{\rm cut}=4000$~fm$^{-1}$. The table shows that the 2D binding energy and
the expectation-value sum $\langle H_0\rangle+\langle V\rangle$ approach stable
limits as both $N_P$ and $N_X$ are increased. In particular, the results become
essentially stable once $N_P$ reaches the range $1536$--$2048$ and
$N_X \gtrsim 128$. For the largest meshes studied, the remaining variation in
$E_2$ is only of order a few $10^{-9}$~MeV, demonstrating that the quoted 2D
results are well converged with respect to the angular discretization. The
noticeable deterioration at $N_X=192$ indicates the onset of numerical
sensitivity in the angular quadrature at very large meshes and does not improve
the final accuracy. On this basis, we adopted $N_P=2048$ and values up to
$N_X=160$ for the final full vector-variable calculations.

For these non-separable potentials, the coordinate-space wave function is
obtained by a numerical Fourier transform of the momentum-space solution. The
accuracy of this step is controlled by the same grid parameters. A
sufficiently large $p_{\rm cut}$ is required to capture the short-range
structure of $\psi(r)$, and a dense low-momentum region is required to
reproduce the long-range asymptotic tail.
Tables~\ref{tab:fourier_rspace_convergence_mt3} and
\ref{tab:fourier_rspace_convergence_mt4} list the coordinate-space
expectation values $\langle H_0\rangle_{\rm cut}$ and $\langle V\rangle_{\rm
cut}$ obtained from the transformed wave functions, evaluated using $N_R$
radial mesh points. For both MT-III and MT-IV, the consistency measure
$\Delta E \equiv (\langle H_0\rangle_{\rm cut}+\langle V\rangle_{\rm
cut})-E_2$ reaches the $10^{-8}$~MeV level once $r_{\rm cut}\gtrsim 40$~fm.
This demonstrates that the numerical Fourier transformation handles the stiff
repulsive core of the MT-IV potential with high precision.
Table~\ref{tab:fourier_rspace_convergence_mt34_radius} reports the
corresponding radii as functions of $r_{\rm cut}$.

\begin{figure*}[p]
    \centering
    \hspace{-0.2cm}
    \includegraphics[width=8cm,trim=0cm 0.1cm 0.15cm 0cm, clip]{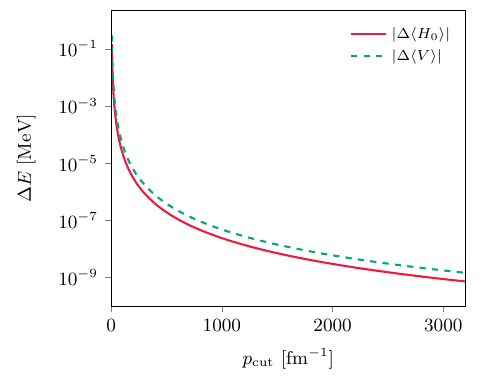}
    \hspace{0.8cm}
    \includegraphics[width=8cm,trim=0cm 0.1cm 0.15cm 0cm, clip]{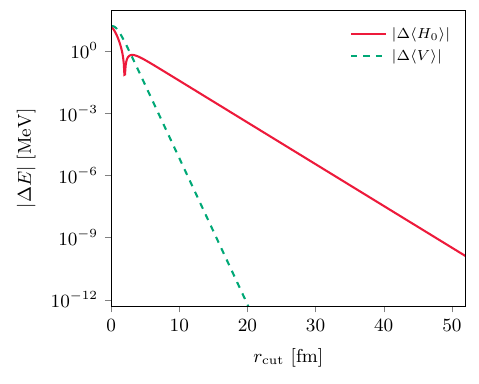}
    \vspace{-0.4cm}
    \caption{\label{fig:cutconvergence}
    \linespread{0.98}\selectfont Convergence of the deviations $\Delta\langle H_0\rangle$ and $\Delta\langle V\rangle$
    from their exact values as functions of the momentum cut-off $p_{\rm cut}$ (left) and
    the radial cut-off $r_{\rm cut}$ (right) for the YAMA-23 potential.}
    \vspace{0.5cm}
    \hspace{-0.2cm}
    \includegraphics[width=8cm,trim=0cm 0.1cm 0.150cm 0cm, clip]{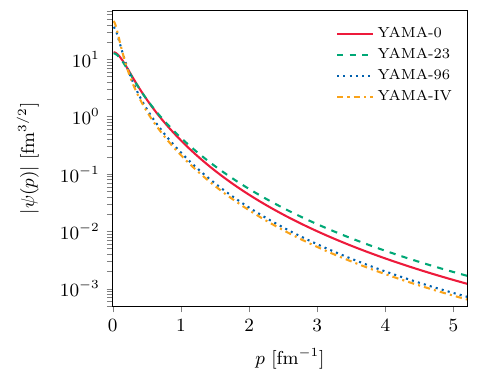}
    \hspace{0.8cm}
    \includegraphics[width=8cm,trim=0cm 0.1cm 0.15cm 0cm, clip]{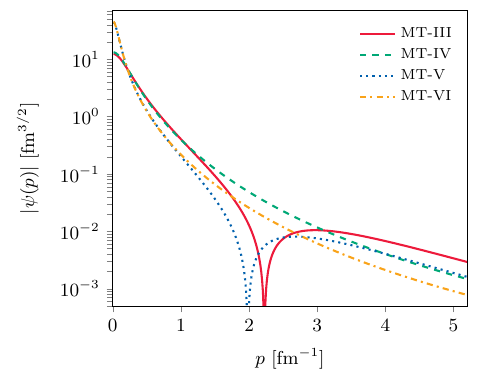}
    \vspace{-0.4cm}
    \caption{\label{fig:wavemomentum}
    \linespread{0.98}\selectfont
    Momentum-space wave functions $\psi(p)$ for selected Yamaguchi (left) and Malfliet--Tjon (right)
    potentials, obtained from the solution of the Lippmann--Schwinger equation for $l=0$.}
    \vspace{0.5cm}
    \hspace{-0.2cm}
    \includegraphics[width=8cm,trim=0cm 0.1cm 0.15cm 0cm, clip]{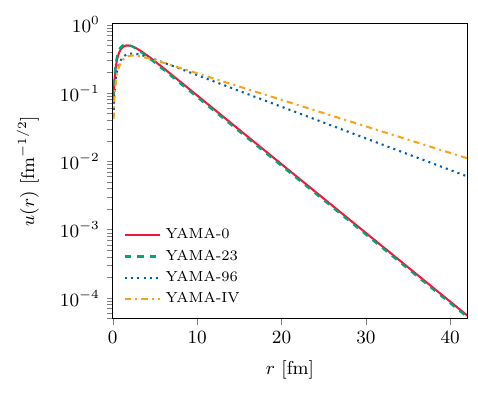}
    \hspace{0.8cm}
    \includegraphics[width=8cm,trim=0cm 0.1cm 0.15cm 0cm, clip]{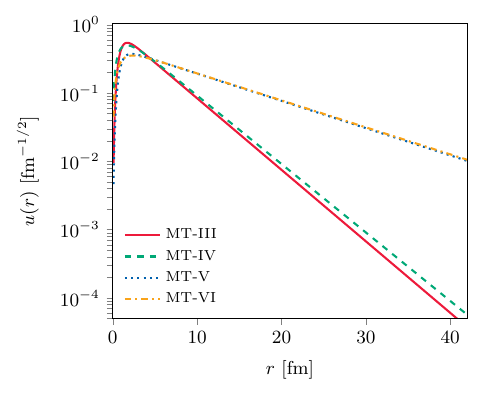}
    \vspace{-0.4cm}
    \caption{\label{fig:waverspace}
    \linespread{0.98}\selectfont
    Reduced coordinate-space wave functions $u(r)$ for selected Yamaguchi (left) and Malfliet--Tjon
    (right) potentials, obtained by Fourier transformation of the momentum-space solutions.}
\end{figure*}

Finally, Table~\ref{tab:yama_mt2d3d} presents a detailed comparison of the
binding energies obtained using the 1D and 2D formulations. We consider three
cases for the 2D approach: restricting the potential expansion to $l=0$,
including partial waves up to $l=12$, and using the full vector-variable
formulation without partial-wave decomposition ($l=\infty$). For the cases
using interactions constructed from partial-wave components ($l=0$ and $l=12$), $N_X = 48$
angular points were sufficient to reach convergence. For the full $l=\infty$
formulation, up to $N_X = 160$ points were used to ensure stability of the
2D angular integration. The computational efficiency of the iterative Arnoldi
solver allowed for the use of the dense $N_P=2048$ grids described above. With
these parameters, the results for the 2D ($l=0$) and 2D ($l=12$) calculations
match the 1D benchmark to the full displayed precision (ten significant digits).
The full vector calculation ($l=\infty$) exhibits minute deviations on the order
of $10^{-9}$~MeV, which are attributable to the numerical discretization of the
angular integral in Eq.~(\ref{eq:vppxx}). As with the Yamaguchi potential, we
verified the angular structure of the solution by projecting the full 2D wave
function onto Legendre polynomials. The analysis confirms that contributions
from $l>0$ components are suppressed to the level of machine precision ($<
10^{-30}$). This strict preservation of the $s$-wave character within the full
vector formalism confirms the numerical consistency of the implementation.
The wave functions in momentum and coordinate space
are shown in Fig.~\ref{fig:wavemomentum} and Fig.~\ref{fig:waverspace},
respectively. Figure~\ref{fig:wavemomentum} clearly illustrates the substantial
high-momentum tail generated by the short-range repulsion of the MT potentials
compared to the soft Yamaguchi interaction. In coordinate space
(Fig.~\ref{fig:waverspace}), all ground-state wave functions exhibit the
expected node-free structure, with the MT potentials showing characteristic
suppression at short distances due to the repulsive core.

\section{Summary and Outlook}
\label{sec:summary}

In this work, we solved the two-boson bound-state problem in momentum space
using two complementary formulations. We used a traditional one-dimensional
partial-wave approach and a two-dimensional approach based on momentum vectors.
By applying both methods to the analytically solvable Yamaguchi potential and to
the local Malfliet--Tjon interaction, we demonstrated the correctness,
stability, and numerical precision of the implementation.

The main achievements of this study are threefold. First, we demonstrated that
the 2D vector-variable formulation yields binding energies identical to the 1D
partial-wave result to ten decimal places ($10^{-10}$~MeV). This confirms that
the vector-variable approach handles the angular integration and the coupling of
the momentum grid correctly, serving as a controlled methodological benchmark
for more demanding applications. Second, for the Yamaguchi potential, we derived
exact analytical expressions that quantify the systematic errors induced by
finite momentum- and coordinate-space cut-offs. Together with the detailed
convergence tables, these results provide quantitative benchmarks for estimating
numerical uncertainties in momentum-space calculations. Third, the application
to the Malfliet--Tjon potentials, including the consistent treatment of higher
partial waves in the 2D formulation, demonstrates that the methods remain
reliable for non-separable, local interactions with strong short-range
components.

The benchmarked two-body setup presented here provides a building block for more
complex few-body calculations. While full validation of non-PWD methods for
three- and four-body collisions ultimately requires testing the high-energy
off-shell $T$-matrix, the numerical insights from the cut-off studies and the
tested 2D vector-momentum formulation will be used in forthcoming papers on
three-body bound-state and scattering problems. The 2D approach closely mirrors
the structure of the Faddeev equations and is therefore well suited for
extending the present analysis to three-body systems. In this sense, the present
study establishes a well-controlled reference standard for testing the
underlying numerical algorithms prior to full three-body calculations.

Finally, separable interactions such as the Yamaguchi potential remain a
valuable tool in few-body physics. In three-body calculations, they reduce the
dimensionality of the Faddeev equations and provide controlled, analytically
tractable test cases. This supports the development and benchmarking of
higher-dimensional numerical approaches that employ realistic local potentials.
The present work therefore provides a high-precision methodological reference
for the further development of numerical studies in few-body systems.

\appendix

\begingroup
\makeatletter
\renewenvironment{widetext}{%
  \par\ignorespaces
  \onecolumngrid
  \vskip10\p@
  \vskip6\p@
  \prep@math@patch
}{%
  \par
  \vskip6\p@
  \vskip8.5\p@
  \twocolumngrid\global\@ignoretrue
  \@endpetrue
}
\makeatother

\begin{widetext}
\section{Analytical formulas for the Yamaguchi potential in momentum space}
\label{sec:limityamaguchi}

To quantify the systematic error introduced by a finite momentum cut-off $p_{\rm
cut}$, we derive exact analytical expressions for the expectation values of
$H_0$ and $V$ as functions of $p_{\rm cut}$. Using the wave function from
Eq.~(\ref{eq:psiyamaguchipspace}), the potential from
Eq.~(\ref{eq:yamaguchipot}), and the form factors from
Eq.~(\ref{eq:yamaguchiform}) for $l=0$, the integrals can be calculated
analytically. For the kinetic energy we obtain

\begin{alignat}{2}
 \langle \psi | H_0 | \psi \rangle & =  \alpha \, \beta \label{eq:h0exact} \\
&= \lim_{p_{\rm cut} \to \infty} \int\limits_{0}^{p_{\rm cut}} \! dp \, p^2
\, \psi^*(p) \, p^2 \,\psi(p)  \\
&= \lim_{p_{\rm cut} \to \infty} \left[ \frac{2\,\alpha \, \beta}{\pi \,
(\alpha -\beta)^3 } \left( \frac{ (\alpha - \beta )(\alpha +\beta )
   \left(
    2 \, \alpha^2 \, \beta^2 + ( \alpha^2 + \beta^2) \, p_{\rm cut}^2
    \right)   p_{\rm cut} }
   {(p_{\rm cut}^2 + \alpha^2 ) (p_{\rm cut}^2 +\beta^2 )} \right.
   \right. \nonumber \\
& \left.   \qquad \qquad \qquad + \left(\alpha^3 + 3\,\alpha \, \beta^2\right)
\tan^{-1}\!\left(\frac{p_{\rm cut}}{\alpha
   }\right)
   \left. - \beta  \left(3\,\alpha^2 +\beta^2\right) \tan^{-1}\!
   \left(\frac{p_{\rm cut}}{\beta }\right)
   \right)  \right] \!,
\label{eq:h0part}
\intertext{and for the potential energy}
   \langle \psi | V | \psi \rangle  &= - \frac{\pi^2\,\alpha \, \lambda}{\beta \,
   (\alpha + \beta)} \label{eq:vexact} \\
  &= \lim_{p_{\rm cut} \to \infty} -4\pi\lambda \left
  [\int\limits_{0}^{p_{\rm cut}} \! dp \,
   p^2 \,\psi(p) \,
   \frac{1}{p^2 + \beta^2} \,  \right]^{2}  \\
   &= \lim_{p_{\rm cut} \to \infty} \left[ -\frac{4 \, \alpha \,
    \lambda}{ (\alpha - \beta )^4 \,
      \beta \, (\alpha +\beta ) \left( p_{\rm cut}^2 + \beta^2 \right)^2} \,
      \Biggl ( \beta \,(\beta^2 -\alpha^2) \, p_{\rm cut}
     \right. \nonumber \\
   & \left. \qquad \qquad \qquad + \left(p_{\rm cut}^2 + \beta^2\right)
   \left(\left(\alpha^2+ \beta^2\right) \tan^{-1}\!\left(\frac{p_{\rm cut}}{\beta }
   \right)-2\,\alpha \, \beta
   \tan ^{-1}\!\left(\frac{p_{\rm cut}}
   {\alpha }\right) \right) \Biggr )^{\!2\,} \right] \!.
   \label{eq:vpart}
  \end{alignat}
The quadratic structure in the last equation reflects the rank-one separable
form of the Yamaguchi interaction. Since the momentum-space wave function does
not change sign, the cut-off-dependent expressions converge monotonically, as
illustrated in Fig.~\ref{fig:cutconvergence}.

\section{Analytical formulas for the Yamaguchi potential in coordinate space}
\label{sec:waverspace}

We now derive analytical expressions for coordinate-space integrals with a
finite radial cut-off $r_{\rm cut}$, which allow a direct assessment of
truncation errors in coordinate space. Using the wave function from
Eq.~(\ref{eq:psirspace}), the relevant limits can be evaluated analytically. For
the kinetic energy we find

\begin{alignat}{2}
 \langle \psi | H_0 | \psi \rangle & =  \alpha \, \beta
\label{eq:h0exact_rspace}\\
&= \lim_{r_{\rm cut} \to \infty}
\int\limits_{0}^{r_{\rm cut}} \! dr \, r^2 \,
\psi^*(r) \left(-\frac{1}{r^2} \frac{\partial}{\partial r} \left( r^2 \frac{\partial}
{\partial r} \right) \right)  \psi(r) \,  \\
 &=  \lim_{r_{\rm cut}  \to \infty} \Biggl[ -\frac{\alpha \, \beta \,
(\alpha + \beta)}{(\alpha - \beta)^2} \Big( \alpha +
\beta   - \alpha \, e^{-2\,\alpha \, r_{\rm cut}} - \beta \,  e^{-2 \, \beta \, r_{\rm cut}}
 +  \frac{2\,(\alpha^2 + \beta^2)}{\alpha + \beta}
( e^{-(\alpha + \beta)\, r_{\rm cut} } -1 ) \Big) \Biggr] ,
\label{eq:h0part_rspace}
\intertext{and for the potential energy}
   \langle \psi | V | \psi \rangle  &= - \frac{\pi^2\,\alpha \, \lambda}{\beta
   (\alpha + \beta) } \label{eq:vexact_rspace} \\
  &= \lim_{r_{\rm cut} \to \infty} -4\pi\lambda
  \left[ \int\limits_{0}^{r_{\rm cut}} \! dr \, r^2 \, g(r) \, \psi(r)
  \right]^{2}  \\
  &= \lim_{r_{\rm cut} \to \infty} \left[ -
  \frac{4\pi^2\,\alpha \, \beta \, \lambda \,(\alpha+\beta)}{(\alpha-\beta)^2}
  \left(
  \frac{1-e^{-(\alpha+\beta) \, r_{\rm cut}}}{\alpha+\beta} -
  \frac{1-e^{-2 \, \beta \, r_{\rm cut}}}{2 \, \beta}  \right)^{\!2\,} \right].
   \label{eq:vpart_rspace}
\end{alignat}

All cut-off-dependent corrections decay exponentially with $r_{\rm cut}$, in
contrast to the algebraic convergence observed in momentum space. In coordinate
space, the Yamaguchi wave function changes sign, leading to partial
cancellations in the integrals. For sufficiently large $r_{\rm cut}$, however,
the convergence again becomes monotonic, as shown in
Fig.~\ref{fig:cutconvergence}.

Closed-form expressions can also be obtained for spatial observables by
introducing a cut-off $r_{\rm cut}$ and evaluating the corresponding limits
analytically, such as for the mean radius of the two-body system
\begin{alignat}{2}
  \langle \psi | \! \left( \frac{1}{2} \,r \right) \!| \psi \rangle  &=  \frac{1}{4}
 \left( \frac{1}{\alpha}  +  \frac{1}{\beta}+ \frac{2}{\alpha + \beta}\right)
 \label{eq:radiusexact}\\
&= \!\lim_{r_{\rm cut} \to \infty} \int\limits_{0}^{r_{\rm cut}} \! dr \, r^2 \,
\psi^*(r)
\left( \frac{1}{2} \,r \right)  \psi(r) \\
&= \!\lim_{r_{\rm cut} \to \infty} \Biggl[\frac{\alpha \, \beta \,(\alpha +\beta )}
{{4 \,  (\alpha -\beta )^2}}
\left(\frac{1 - (1+ 2\,\alpha \,r_{\rm cut}) \, e^{-2\,\alpha \, r_{\rm cut}} }
{\alpha^2} +\frac{1- (1+ 2 \, \beta \, r_{\rm cut}) \, e^{-2 \, \beta \, r_{\rm cut}}
}{\beta^2}
\right . \nonumber \\
& \qquad \qquad \qquad  \left . -\frac{8-8\,(1+  (\alpha +\beta ) \, r_{\rm cut} )
 \, e^{- (\alpha +\beta ) \, r_{\rm cut}}  }{(\alpha +\beta )^2}\right) \Biggr]\,,
  \label{eq:limitradius}
\intertext{and the rms radius}
   \sqrt{   \langle \psi | \! \left(\frac{1}{2} \,r\right)^{\!2}  \!| \psi \rangle }
  & = \sqrt{\frac{1}{8}
\left(\frac{1}{\alpha^2}+\frac{1}{\beta^2}+\frac{3}{\alpha \, \beta}+
\frac{4}{(\alpha +\beta )^2} \right)}
\label{eq:rmsradiusexact} \\
&= \lim_{r_{\rm cut} \to \infty} \left[ \int\limits_{0}^{r_{\rm cut}} \! dr \, r^2 \,
\psi^*(r) \,
\left(\frac{1}{2} \,r\right)^{\!2} \psi(r) \right]^{ \frac{1}{2}} \\
&= \lim_{r_{\rm cut} \to \infty} \left[ \frac{\alpha \, \beta \, (\alpha +\beta )}
{8 \,(\alpha -\beta )^2} \left(\frac{1}{\alpha^3}-
  \frac{
\left(1+ 2\,\alpha \,r_{\rm cut} \,(1+ \alpha \,r_{\rm cut})\right) e^{-2\,\alpha
\, r_{\rm cut}}}
{\alpha^3} \right. \right. \nonumber \\
& \qquad \qquad \qquad+ \frac{1}{\beta^3}
-     \frac{
\left(1+ 2 \, \beta \, r_{\rm cut} \, (1+ \beta \, r_{\rm cut})\right) e^{-2 \, \beta
 \, r_{\rm cut}}} {\beta^3} - \frac{16}{(\alpha + \beta)^3} \nonumber \\
&  \qquad \qquad \qquad  \left.
   +\frac{8\left(2+ (\alpha +\beta ) \, r_{\rm cut} \,
   ( 2 + (\alpha +\beta ) \, r_{\rm cut}
   )\right)  e^{- (\alpha +\beta ) \, r_{\rm cut}}}{(\alpha +\beta )^3} \right)
 \left. \bigg)\right]^{\! \frac{1}{2}} .
  \label{eq:limitrmsradius}
\end{alignat}
\end{widetext}
\endgroup


\begin{thebibliography}{36}%
\makeatletter
\providecommand \@ifxundefined [1]{%
 \@ifx{#1\undefined}
}%
\providecommand \@ifnum [1]{%
 \ifnum #1\expandafter \@firstoftwo
 \else \expandafter \@secondoftwo
 \fi
}%
\providecommand \@ifx [1]{%
 \ifx #1\expandafter \@firstoftwo
 \else \expandafter \@secondoftwo
 \fi
}%
\providecommand \natexlab [1]{#1}%
\providecommand \enquote  [1]{``#1''}%
\providecommand \bibnamefont  [1]{#1}%
\providecommand \bibfnamefont [1]{#1}%
\providecommand \citenamefont [1]{#1}%
\providecommand \href@noop [0]{\@secondoftwo}%
\providecommand \href [0]{\begingroup \@sanitize@url \@href}%
\providecommand \@href[1]{\@@startlink{#1}\@@href}%
\providecommand \@@href[1]{\endgroup#1\@@endlink}%
\providecommand \@sanitize@url [0]{\catcode `\\12\catcode `\$12\catcode
  `\&12\catcode `\#12\catcode `\^12\catcode `\_12\catcode `\%12\relax}%
\providecommand \@@startlink[1]{}%
\providecommand \@@endlink[0]{}%
\providecommand \url  [0]{\begingroup\@sanitize@url \@url }%
\providecommand \@url [1]{\endgroup\@href {#1}{\urlprefix }}%
\providecommand \urlprefix  [0]{URL }%
\providecommand \Eprint [0]{\href }%
\providecommand \doibase [0]{https://doi.org/}%
\providecommand \selectlanguage [0]{\@gobble}%
\providecommand \bibinfo  [0]{\@secondoftwo}%
\providecommand \bibfield  [0]{\@secondoftwo}%
\providecommand \translation [1]{[#1]}%
\providecommand \BibitemOpen [0]{}%
\providecommand \bibitemStop [0]{}%
\providecommand \bibitemNoStop [0]{.\EOS\space}%
\providecommand \EOS [0]{\spacefactor3000\relax}%
\providecommand \BibitemShut  [1]{\csname bibitem#1\endcsname}%
\let\auto@bib@innerbib\@empty
\bibitem [{\citenamefont {{C}h. Elster}\ \emph {et~al.}(1998)\citenamefont
  {{C}h. Elster}, \citenamefont {Thomas},\ and\ \citenamefont
  {Gl{\"o}ckle}}]{Elster98a}%
  \BibitemOpen
  \bibfield  {author} {\bibinfo {author} {\bibnamefont {{C}h. Elster}},
  \bibinfo {author} {\bibfnamefont {J.~H.}\ \bibnamefont {Thomas}},\ and\
  \bibinfo {author} {\bibfnamefont {W.}~\bibnamefont {Gl{\"o}ckle}},\ }\href
  {https://doi.org/10.1007/s006010050076} {\bibfield  {journal} {\bibinfo
  {journal} {Few-Body Systems}\ }\textbf {\bibinfo {volume} {24}},\ \bibinfo
  {pages} {55} (\bibinfo {year} {1998})}\BibitemShut {NoStop}%
\bibitem [{\citenamefont {Elster}\ \emph {et~al.}(1999)\citenamefont {Elster},
  \citenamefont {Schadow}, \citenamefont {Nogga},\ and\ \citenamefont
  {Gl{\"o}ckle}}]{Elster98b}%
  \BibitemOpen
  \bibfield  {author} {\bibinfo {author} {\bibfnamefont {{\relax
  Ch}.}~\bibnamefont {Elster}}, \bibinfo {author} {\bibfnamefont
  {W.}~\bibnamefont {Schadow}}, \bibinfo {author} {\bibfnamefont
  {A.}~\bibnamefont {Nogga}},\ and\ \bibinfo {author} {\bibfnamefont
  {W.}~\bibnamefont {Gl{\"o}ckle}},\ }\href
  {https://doi.org/10.1007/s006010050124} {\bibfield  {journal} {\bibinfo
  {journal} {Few-Body Systems}\ }\textbf {\bibinfo {volume} {27}},\ \bibinfo
  {pages} {83} (\bibinfo {year} {1999})}\BibitemShut {NoStop}%
\bibitem [{\citenamefont {Schadow}\ \emph {et~al.}(2000)\citenamefont
  {Schadow}, \citenamefont {Elster},\ and\ \citenamefont
  {Gl{\"o}ckle}}]{Schadow99d}%
  \BibitemOpen
  \bibfield  {author} {\bibinfo {author} {\bibfnamefont {W.}~\bibnamefont
  {Schadow}}, \bibinfo {author} {\bibfnamefont {{\relax Ch}.}~\bibnamefont
  {Elster}},\ and\ \bibinfo {author} {\bibfnamefont {W.}~\bibnamefont
  {Gl{\"o}ckle}},\ }\href {https://doi.org/10.1007/s006010070028} {\bibfield
  {journal} {\bibinfo  {journal} {Few-Body Systems}\ }\textbf {\bibinfo
  {volume} {28}},\ \bibinfo {pages} {15} (\bibinfo {year} {2000})}\BibitemShut
  {NoStop}%
\bibitem [{\citenamefont {{M.~Pol{\'a}{\v{s}}ek and M.~Ingr and
  P.~{\v{C}}{\'a}rsky and J.~Hor{\'a}{\v{c}}ek}}(2000)}]{Polasek2000}%
  \BibitemOpen
  \bibfield  {author} {\bibinfo {author} {\bibnamefont {{M.~Pol{\'a}{\v{s}}ek
  and M.~Ingr and P.~{\v{C}}{\'a}rsky and J.~Hor{\'a}{\v{c}}ek}}},\ }\href
  {https://doi.org/10.1103/PhysRevA.61.032701} {\bibfield  {journal} {\bibinfo
  {journal} {Phys. Rev. A}\ }\textbf {\bibinfo {volume} {61}},\ \bibinfo
  {pages} {032701} (\bibinfo {year} {2000})}\BibitemShut {NoStop}%
\bibitem [{\citenamefont {{J.~Shertzer and A.~Temkin}}(2001)}]{Shertzer2001}%
  \BibitemOpen
  \bibfield  {author} {\bibinfo {author} {\bibnamefont {{J.~Shertzer and
  A.~Temkin}}},\ }\href {https://doi.org/10.1103/PhysRevA.63.062714} {\bibfield
   {journal} {\bibinfo  {journal} {Phys. Rev. A}\ }\textbf {\bibinfo {volume}
  {63}},\ \bibinfo {pages} {062714} (\bibinfo {year} {2001})}\BibitemShut
  {NoStop}%
\bibitem [{\citenamefont {{G.~L.~Caia and V.~Pascalutsa and
  L.~E.~Wright}}(2004)}]{Caia2004}%
  \BibitemOpen
  \bibfield  {author} {\bibinfo {author} {\bibnamefont {{G.~L.~Caia and
  V.~Pascalutsa and L.~E.~Wright}}},\ }\href
  {https://doi.org/10.1103/PhysRevC.69.034003} {\bibfield  {journal} {\bibinfo
  {journal} {Phys. Rev. C}\ }\textbf {\bibinfo {volume} {69}},\ \bibinfo
  {pages} {034003} (\bibinfo {year} {2004})}\BibitemShut {NoStop}%
\bibitem [{\citenamefont {{B.~Kessler and G.~L.~Payne and
  W.~N.~Polyzou}}(2004)}]{Kessler2004}%
  \BibitemOpen
  \bibfield  {author} {\bibinfo {author} {\bibnamefont {{B.~Kessler and
  G.~L.~Payne and W.~N.~Polyzou}}},\ }\href
  {https://doi.org/10.1103/PhysRevC.70.034003} {\bibfield  {journal} {\bibinfo
  {journal} {Phys. Rev. C}\ }\textbf {\bibinfo {volume} {70}},\ \bibinfo
  {pages} {034003} (\bibinfo {year} {2004})}\BibitemShut {NoStop}%
\bibitem [{\citenamefont {{A.~S.~Kadyrov and I.~Bray and A.~T.~Stelbovics and
  B.~Saha}}(2005)}]{Kadyrov2005}%
  \BibitemOpen
  \bibfield  {author} {\bibinfo {author} {\bibnamefont {{A.~S.~Kadyrov and
  I.~Bray and A.~T.~Stelbovics and B.~Saha}}},\ }\href
  {https://doi.org/10.1088/0953-4075/38/5/004} {\bibfield  {journal} {\bibinfo
  {journal} {J. Phys. B: At. Mol. Opt. Phys.}\ }\textbf {\bibinfo {volume}
  {38}},\ \bibinfo {pages} {509} (\bibinfo {year} {2005})}\BibitemShut
  {NoStop}%
\bibitem [{\citenamefont {Liu}\ \emph {et~al.}(2005)\citenamefont {Liu},
  \citenamefont {Elster},\ and\ \citenamefont {Glöckle}}]{Liu2005a}%
  \BibitemOpen
  \bibfield  {author} {\bibinfo {author} {\bibfnamefont {H.}~\bibnamefont
  {Liu}}, \bibinfo {author} {\bibfnamefont {{\relax Ch}.}~\bibnamefont
  {Elster}},\ and\ \bibinfo {author} {\bibfnamefont {W.}~\bibnamefont
  {Glöckle}},\ }\href {https://doi.org/10.1103/PhysRevC.72.054003} {\bibfield
  {journal} {\bibinfo  {journal} {Phys. Rev. C}\ }\textbf {\bibinfo {volume}
  {72}},\ \bibinfo {pages} {054003} (\bibinfo {year} {2005})}\BibitemShut
  {NoStop}%
\bibitem [{\citenamefont {{G.~Ramalho and A.~Arriaga and
  M.~T.~Pe{\~n}a}}(2006)}]{Ramalho2006}%
  \BibitemOpen
  \bibfield  {author} {\bibinfo {author} {\bibnamefont {{G.~Ramalho and
  A.~Arriaga and M.~T.~Pe{\~n}a}}},\ }\href
  {https://doi.org/10.1007/s00601-006-0161-3} {\bibfield  {journal} {\bibinfo
  {journal} {Few-Body Systems}\ }\textbf {\bibinfo {volume} {39}},\ \bibinfo
  {pages} {123} (\bibinfo {year} {2006})}\BibitemShut {NoStop}%
\bibitem [{\citenamefont {Liu}\ \emph {et~al.}(2007)\citenamefont {Liu},
  \citenamefont {Elster},\ and\ \citenamefont {Glöckle}}]{Liu2007a}%
  \BibitemOpen
  \bibfield  {author} {\bibinfo {author} {\bibfnamefont {H.}~\bibnamefont
  {Liu}}, \bibinfo {author} {\bibfnamefont {{\relax Ch}.}~\bibnamefont
  {Elster}},\ and\ \bibinfo {author} {\bibfnamefont {W.}~\bibnamefont
  {Glöckle}},\ }\href {https://doi.org/10.1016/j.nuclphysa.2007.03.147}
  {\bibfield  {journal} {\bibinfo  {journal} {Nucl. Phys.}\ }\textbf {\bibinfo
  {volume} {A790}},\ \bibinfo {pages} {262c} (\bibinfo {year}
  {2007})}\BibitemShut {NoStop}%
\bibitem [{\citenamefont {Hadizadeh}\ and\ \citenamefont
  {Bayegan}(2007)}]{Hadizadeh2007a}%
  \BibitemOpen
  \bibfield  {author} {\bibinfo {author} {\bibfnamefont {M.~R.}\ \bibnamefont
  {Hadizadeh}}\ and\ \bibinfo {author} {\bibfnamefont {S.}~\bibnamefont
  {Bayegan}},\ }\href {https://doi.org/10.1007/s00601-006-0169-8} {\bibfield
  {journal} {\bibinfo  {journal} {Few-Body Systems}\ }\textbf {\bibinfo
  {volume} {40}},\ \bibinfo {pages} {171} (\bibinfo {year} {2007})}\BibitemShut
  {NoStop}%
\bibitem [{\citenamefont {{M.~Rodriguez-Gallardo and A.~Deltuva and E.~Cravo
  and R.~Crespo and A.~C.~Fonseca}}(2008)}]{RodriguezGallardo2008}%
  \BibitemOpen
  \bibfield  {author} {\bibinfo {author} {\bibnamefont {{M.~Rodriguez-Gallardo
  and A.~Deltuva and E.~Cravo and R.~Crespo and A.~C.~Fonseca}}},\ }\href
  {https://doi.org/10.1103/PhysRevC.78.034602} {\bibfield  {journal} {\bibinfo
  {journal} {Phys. Rev. C}\ }\textbf {\bibinfo {volume} {78}},\ \bibinfo
  {pages} {034602} (\bibinfo {year} {2008})}\BibitemShut {NoStop}%
\bibitem [{\citenamefont {Hadizadeh}\ and\ \citenamefont
  {Bayegan}(2008)}]{Hadizadeh2008a}%
  \BibitemOpen
  \bibfield  {author} {\bibinfo {author} {\bibfnamefont {M.~R.}\ \bibnamefont
  {Hadizadeh}}\ and\ \bibinfo {author} {\bibfnamefont {S.}~\bibnamefont
  {Bayegan}},\ }\href {https://doi.org/10.1140/epja/i2008-10583-8} {\bibfield
  {journal} {\bibinfo  {journal} {Eur. Phys. J. A}\ }\textbf {\bibinfo {volume}
  {36}},\ \bibinfo {pages} {201} (\bibinfo {year} {2008})}\BibitemShut
  {NoStop}%
\bibitem [{\citenamefont {Bayegan}\ \emph {et~al.}(2008)\citenamefont
  {Bayegan}, \citenamefont {Hadizadeh},\ and\ \citenamefont
  {Harzchi}}]{Bayegan2008b}%
  \BibitemOpen
  \bibfield  {author} {\bibinfo {author} {\bibfnamefont {S.}~\bibnamefont
  {Bayegan}}, \bibinfo {author} {\bibfnamefont {M.~R.}\ \bibnamefont
  {Hadizadeh}},\ and\ \bibinfo {author} {\bibfnamefont {M.}~\bibnamefont
  {Harzchi}},\ }\href {https://doi.org/10.1103/PhysRevC.77.064005} {\bibfield
  {journal} {\bibinfo  {journal} {Phys. Rev. C}\ }\textbf {\bibinfo {volume}
  {77}},\ \bibinfo {pages} {064005} (\bibinfo {year} {2008})}\BibitemShut
  {NoStop}%
\bibitem [{\citenamefont {Harzchi}\ and\ \citenamefont
  {Bayegan}(2010)}]{Harzchi2010a}%
  \BibitemOpen
  \bibfield  {author} {\bibinfo {author} {\bibfnamefont {M.}~\bibnamefont
  {Harzchi}}\ and\ \bibinfo {author} {\bibfnamefont {S.}~\bibnamefont
  {Bayegan}},\ }\href {https://doi.org/10.1140/epja/i2010-11039-4} {\bibfield
  {journal} {\bibinfo  {journal} {Eur. Phys. J.}\ }\textbf {\bibinfo {volume}
  {A46}},\ \bibinfo {pages} {271} (\bibinfo {year} {2010})}\BibitemShut
  {NoStop}%
\bibitem [{\citenamefont {Gl{\"o}ckle}\ \emph {et~al.}(2010)\citenamefont
  {Gl{\"o}ckle}, \citenamefont {Elster}, \citenamefont {Golak}, \citenamefont
  {Skibi{\'n}ski}, \citenamefont {Wita{\l}a},\ and\ \citenamefont
  {Kamada}}]{Gloeckle2010a}%
  \BibitemOpen
  \bibfield  {author} {\bibinfo {author} {\bibfnamefont {W.}~\bibnamefont
  {Gl{\"o}ckle}}, \bibinfo {author} {\bibfnamefont {{\relax Ch}.}~\bibnamefont
  {Elster}}, \bibinfo {author} {\bibfnamefont {J.}~\bibnamefont {Golak}},
  \bibinfo {author} {\bibfnamefont {R.}~\bibnamefont {Skibi{\'n}ski}}, \bibinfo
  {author} {\bibfnamefont {H.}~\bibnamefont {Wita{\l}a}},\ and\ \bibinfo
  {author} {\bibfnamefont {H.}~\bibnamefont {Kamada}},\ }\href
  {https://doi.org/10.1007/s00601-009-0064-1} {\bibfield  {journal} {\bibinfo
  {journal} {Few-Body Systems}\ }\textbf {\bibinfo {volume} {47}},\ \bibinfo
  {pages} {25} (\bibinfo {year} {2010})}\BibitemShut {NoStop}%
\bibitem [{\citenamefont {Golak}\ \emph {et~al.}(2010)\citenamefont {Golak},
  \citenamefont {Gl\"ockle}, \citenamefont {Skibi{\'{n}}ski}, \citenamefont
  {Wita{\l}a}, \citenamefont {Rozpedzik}, \citenamefont {Topolnicki},
  \citenamefont {Fachruddin}, \citenamefont {Elster},\ and\ \citenamefont
  {Nogga}}]{Golak2010a}%
  \BibitemOpen
  \bibfield  {author} {\bibinfo {author} {\bibfnamefont {J.}~\bibnamefont
  {Golak}}, \bibinfo {author} {\bibfnamefont {W.}~\bibnamefont {Gl\"ockle}},
  \bibinfo {author} {\bibfnamefont {R.}~\bibnamefont {Skibi{\'{n}}ski}},
  \bibinfo {author} {\bibfnamefont {H.}~\bibnamefont {Wita{\l}a}}, \bibinfo
  {author} {\bibfnamefont {D.}~\bibnamefont {Rozpedzik}}, \bibinfo {author}
  {\bibfnamefont {K.}~\bibnamefont {Topolnicki}}, \bibinfo {author}
  {\bibfnamefont {I.}~\bibnamefont {Fachruddin}}, \bibinfo {author}
  {\bibfnamefont {{\relax Ch}.}~\bibnamefont {Elster}},\ and\ \bibinfo {author}
  {\bibfnamefont {A.}~\bibnamefont {Nogga}},\ }\href
  {https://doi.org/10.1103/PhysRevC.81.034006} {\bibfield  {journal} {\bibinfo
  {journal} {Phys. Rev. C}\ }\textbf {\bibinfo {volume} {81}},\ \bibinfo
  {pages} {034006} (\bibinfo {year} {2010})}\BibitemShut {NoStop}%
\bibitem [{\citenamefont {Golak}\ \emph {et~al.}(2012)\citenamefont {Golak},
  \citenamefont {Topolnicki}, \citenamefont {Skibiński}, \citenamefont
  {Glöckle}, \citenamefont {Kamada},\ and\ \citenamefont
  {Nogga}}]{Golak2012a}%
  \BibitemOpen
  \bibfield  {author} {\bibinfo {author} {\bibfnamefont {J.}~\bibnamefont
  {Golak}}, \bibinfo {author} {\bibfnamefont {K.}~\bibnamefont {Topolnicki}},
  \bibinfo {author} {\bibfnamefont {R.}~\bibnamefont {Skibiński}}, \bibinfo
  {author} {\bibfnamefont {W.}~\bibnamefont {Glöckle}}, \bibinfo {author}
  {\bibfnamefont {H.}~\bibnamefont {Kamada}},\ and\ \bibinfo {author}
  {\bibfnamefont {A.}~\bibnamefont {Nogga}},\ }\href
  {https://doi.org/10.1007/s00601-012-0472-5} {\bibfield  {journal} {\bibinfo
  {journal} {Few-Body Systems}\ }\textbf {\bibinfo {volume} {54}},\ \bibinfo
  {pages} {2427} (\bibinfo {year} {2012})}\BibitemShut {NoStop}%
\bibitem [{\citenamefont {Shalchi}\ and\ \citenamefont
  {Bayegan}(2012)}]{Shalchi2012a}%
  \BibitemOpen
  \bibfield  {author} {\bibinfo {author} {\bibfnamefont {M.~A.}\ \bibnamefont
  {Shalchi}}\ and\ \bibinfo {author} {\bibfnamefont {S.}~\bibnamefont
  {Bayegan}},\ }\href {https://doi.org/10.1140/epja/i2012-12006-9} {\bibfield
  {journal} {\bibinfo  {journal} {Eur. Phys. J.}\ }\textbf {\bibinfo {volume}
  {A48}},\ \bibinfo {pages} {6} (\bibinfo {year} {2012})}\BibitemShut {NoStop}%
\bibitem [{\citenamefont {{S.~Veerasamy and {C}h.~Elster and
  W.~N.~Polyzou}}(2013)}]{Veerasamy2013}%
  \BibitemOpen
  \bibfield  {author} {\bibinfo {author} {\bibnamefont {{S.~Veerasamy and
  {C}h.~Elster and W.~N.~Polyzou}}},\ }\href
  {https://doi.org/10.1007/s00601-012-0476-1} {\bibfield  {journal} {\bibinfo
  {journal} {Few-Body Systems}\ }\textbf {\bibinfo {volume} {54}},\ \bibinfo
  {pages} {2207} (\bibinfo {year} {2013})}\BibitemShut {NoStop}%
\bibitem [{\citenamefont {Harzchi}\ and\ \citenamefont
  {Bayegan}(2014)}]{Harzchi2014a}%
  \BibitemOpen
  \bibfield  {author} {\bibinfo {author} {\bibfnamefont {M.}~\bibnamefont
  {Harzchi}}\ and\ \bibinfo {author} {\bibfnamefont {S.}~\bibnamefont
  {Bayegan}},\ }\href {https://doi.org/10.1007/s40094-014-0112-1} {\bibfield
  {journal} {\bibinfo  {journal} {J. Theor. Appl. Phys.}\ }\textbf {\bibinfo
  {volume} {8}},\ \bibinfo {pages} {112} (\bibinfo {year} {2014})}\BibitemShut
  {NoStop}%
\bibitem [{\citenamefont {{Z.~C.~Kuruo{\u{g}}lu}}(2016)}]{Kuruoglu2016}%
  \BibitemOpen
  \bibfield  {author} {\bibinfo {author} {\bibnamefont
  {{Z.~C.~Kuruo{\u{g}}lu}}},\ }\href
  {https://doi.org/10.1103/PhysRevE.94.053303} {\bibfield  {journal} {\bibinfo
  {journal} {Phys. Rev. E}\ }\textbf {\bibinfo {volume} {94}},\ \bibinfo
  {pages} {053308} (\bibinfo {year} {2016})}\BibitemShut {NoStop}%
\bibitem [{\citenamefont {Yamaguchi}(1954{\natexlab{a}})}]{Yamaguchi54a}%
  \BibitemOpen
  \bibfield  {author} {\bibinfo {author} {\bibfnamefont {Y.}~\bibnamefont
  {Yamaguchi}},\ }\href {https://doi.org/10.1103/PhysRev.95.1628} {\bibfield
  {journal} {\bibinfo  {journal} {Phys. Rev.}\ }\textbf {\bibinfo {volume}
  {95}},\ \bibinfo {pages} {1628} (\bibinfo {year}
  {1954}{\natexlab{a}})}\BibitemShut {NoStop}%
\bibitem [{\citenamefont {Yamaguchi}(1954{\natexlab{b}})}]{Yamaguchi54b}%
  \BibitemOpen
  \bibfield  {author} {\bibinfo {author} {\bibfnamefont {Y.}~\bibnamefont
  {Yamaguchi}},\ }\href {https://doi.org/10.1103/PhysRev.95.1635} {\bibfield
  {journal} {\bibinfo  {journal} {Phys. Rev.}\ }\textbf {\bibinfo {volume}
  {95}},\ \bibinfo {pages} {1635} (\bibinfo {year}
  {1954}{\natexlab{b}})}\BibitemShut {NoStop}%
\bibitem [{\citenamefont {Malfliet}\ and\ \citenamefont
  {Tjon}(1969)}]{Malfliet69}%
  \BibitemOpen
  \bibfield  {author} {\bibinfo {author} {\bibfnamefont {R.~A.}\ \bibnamefont
  {Malfliet}}\ and\ \bibinfo {author} {\bibfnamefont {J.~A.}\ \bibnamefont
  {Tjon}},\ }\href {https://doi.org/10.1016/0375-9474(69)90775-1} {\bibfield
  {journal} {\bibinfo  {journal} {Nucl. Phys.}\ }\textbf {\bibinfo {volume}
  {A127}},\ \bibinfo {pages} {161} (\bibinfo {year} {1969})}\BibitemShut
  {NoStop}%
\bibitem [{\citenamefont {Yukawa}(1935)}]{Yukawa35a}%
  \BibitemOpen
  \bibfield  {author} {\bibinfo {author} {\bibfnamefont {H.}~\bibnamefont
  {Yukawa}},\ }\href {https://doi.org/10.11429/ppmsj1919.17.0_48} {\bibfield
  {journal} {\bibinfo  {journal} {Proc. Phys.-Math. Soc. (Japan)}\ }\textbf
  {\bibinfo {volume} {17}},\ \bibinfo {pages} {48} (\bibinfo {year}
  {1935})}\BibitemShut {NoStop}%
\bibitem [{\citenamefont {Mohammadi~Sabet}(2021)}]{Sabet2021}%
  \BibitemOpen
  \bibfield  {author} {\bibinfo {author} {\bibfnamefont {M.}~\bibnamefont
  {Mohammadi~Sabet}},\ }\href {https://doi.org/10.12693/APhysPolA.140.97}
  {\bibfield  {journal} {\bibinfo  {journal} {Acta Phys. Pol. A}\ }\textbf
  {\bibinfo {volume} {140}},\ \bibinfo {pages} {97} (\bibinfo {year}
  {2021})}\BibitemShut {NoStop}%
\bibitem [{\citenamefont {Hamzavi}\ \emph {et~al.}(2012)\citenamefont
  {Hamzavi}, \citenamefont {Movahedi}, \citenamefont {Thylwe},\ and\
  \citenamefont {Rajabi}}]{Hamzavi2012}%
  \BibitemOpen
  \bibfield  {author} {\bibinfo {author} {\bibfnamefont {M.}~\bibnamefont
  {Hamzavi}}, \bibinfo {author} {\bibfnamefont {M.}~\bibnamefont {Movahedi}},
  \bibinfo {author} {\bibfnamefont {K.-E.}\ \bibnamefont {Thylwe}},\ and\
  \bibinfo {author} {\bibfnamefont {A.~A.}\ \bibnamefont {Rajabi}},\ }\href
  {https://doi.org/10.1088/0256-307X/29/8/080302} {\bibfield  {journal}
  {\bibinfo  {journal} {Chin. Phys. Lett.}\ }\textbf {\bibinfo {volume} {29}},\
  \bibinfo {pages} {080302} (\bibinfo {year} {2012})}\BibitemShut {NoStop}%
\bibitem [{\citenamefont {Napsuciale}\ and\ \citenamefont
  {Rodr\'{\i}guez}(2021)}]{Napsuciale2021}%
  \BibitemOpen
  \bibfield  {author} {\bibinfo {author} {\bibfnamefont {M.}~\bibnamefont
  {Napsuciale}}\ and\ \bibinfo {author} {\bibfnamefont {S.}~\bibnamefont
  {Rodr\'{\i}guez}},\ }\href {https://doi.org/10.1016/j.physletb.2021.136218}
  {\bibfield  {journal} {\bibinfo  {journal} {Physics Letters B}\ }\textbf
  {\bibinfo {volume} {816}},\ \bibinfo {pages} {136218} (\bibinfo {year}
  {2021})}\BibitemShut {NoStop}%
\bibitem [{\citenamefont {Gl{\"ockle}}(1991)}]{Gloeckle91}%
  \BibitemOpen
  \bibfield  {author} {\bibinfo {author} {\bibfnamefont {W.}~\bibnamefont
  {Gl{\"ockle}}},\ }\href@noop {} {\emph {\bibinfo {title} {Computational
  Nuclear Physics I.~Nuclear Structure, Edited by K.~Langanke, J.~A.~Maruhn,
  and S.~E.~Koonin, pp 152}}}\ (\bibinfo  {publisher} {Springer-Verlag},\
  \bibinfo {address} {Berlin-Heidelberg},\ \bibinfo {year} {1991})\BibitemShut
  {NoStop}%
\bibitem [{\citenamefont {{W.~A.~Karr and C.~R.~Jamell and
  Y.~N.~Joglekar}}(2010)}]{Karr2010}%
  \BibitemOpen
  \bibfield  {author} {\bibinfo {author} {\bibnamefont {{W.~A.~Karr and
  C.~R.~Jamell and Y.~N.~Joglekar}}},\ }\href
  {https://doi.org/10.1119/1.3272021} {\bibfield  {journal} {\bibinfo
  {journal} {Am. J. Phys.}\ }\textbf {\bibinfo {volume} {78}},\ \bibinfo
  {pages} {407} (\bibinfo {year} {2010})}\BibitemShut {NoStop}%
\bibitem [{\citenamefont {Wilkinson}(1965)}]{Wilkinson65}%
  \BibitemOpen
  \bibfield  {author} {\bibinfo {author} {\bibfnamefont {J.~H.}\ \bibnamefont
  {Wilkinson}},\ }\href@noop {} {\emph {\bibinfo {title} {The Algebraic
  Eigenvalue Problem}}}\ (\bibinfo  {publisher} {Clarendon Press},\ \bibinfo
  {address} {Oxford},\ \bibinfo {year} {1965})\BibitemShut {NoStop}%
\bibitem [{\citenamefont {Gl{\"o}ckle}(1982)}]{Gloeckle82b}%
  \BibitemOpen
  \bibfield  {author} {\bibinfo {author} {\bibfnamefont {W.}~\bibnamefont
  {Gl{\"o}ckle}},\ }\href {https://doi.org/10.1016/0375-9474(82)90364-5}
  {\bibfield  {journal} {\bibinfo  {journal} {Nucl. Phys.}\ }\textbf {\bibinfo
  {volume} {A381}},\ \bibinfo {pages} {343} (\bibinfo {year}
  {1982})}\BibitemShut {NoStop}%
\bibitem [{\citenamefont {Arnoldi}(1951)}]{Arnoldi51}%
  \BibitemOpen
  \bibfield  {author} {\bibinfo {author} {\bibfnamefont {W.~E.}\ \bibnamefont
  {Arnoldi}},\ }\href {https://doi.org/10.1090/qam/42792} {\bibfield  {journal}
  {\bibinfo  {journal} {Q. Appl. Math.}\ }\textbf {\bibinfo {volume} {9}},\
  \bibinfo {pages} {17} (\bibinfo {year} {1951})}\BibitemShut {NoStop}%
\bibitem [{\citenamefont {Stadler}\ \emph {et~al.}(1991)\citenamefont
  {Stadler}, \citenamefont {Gl{\"o}ckle},\ and\ \citenamefont
  {Sauer}}]{Stadler91a}%
  \BibitemOpen
  \bibfield  {author} {\bibinfo {author} {\bibfnamefont {A.}~\bibnamefont
  {Stadler}}, \bibinfo {author} {\bibfnamefont {W.}~\bibnamefont
  {Gl{\"o}ckle}},\ and\ \bibinfo {author} {\bibfnamefont {P.~U.}\ \bibnamefont
  {Sauer}},\ }\href {https://doi.org/10.1103/PhysRevC.44.2319} {\bibfield
  {journal} {\bibinfo  {journal} {Phys. Rev. C}\ }\textbf {\bibinfo {volume}
  {44}},\ \bibinfo {pages} {2319} (\bibinfo {year} {1991})}\BibitemShut
  {NoStop}%
\end{thebibliography}


%

\end{document}